\documentclass[aps,prb,superscriptaddress,reprint]{revtex4-2}
\usepackage[T1]{fontenc}
\usepackage{newtxtext}
\usepackage{newtxmath}
\usepackage{amsmath}
\usepackage{physics2}
\usephysicsmodule{ab,ab.legacy,braket,nabla.legacy,op.legacy}
\usepackage{diffcoeff}
\usepackage{mathtools}
\usepackage{siunitx}
\usepackage{comment}
\usepackage{graphicx}
\usepackage[dvipsnames]{xcolor}
\usepackage{hyperref}
\hypersetup{colorlinks=true,linkcolor=Blue,citecolor=Blue,urlcolor=Blue}
\usepackage[version=4]{mhchem}
\usepackage{bm}

\begin{document}
\title{Effective electron coupling to phonon mechanical angular momentum in helical systems}
\author{Akihito Kato}
\affiliation{Department of Physics and Electronics, Osaka Metropolitan University, 1-1 Naka-ku, Sakai, Osaka 599-8531, Japan}
\affiliation{Institute for Molecular Science, National Institutes of Natural Sciences, Okazaki 444-8585, Japan}
\author{Nobuhiko Yokoshi}
\affiliation{Department of Physics and Electronics, Osaka Metropolitan University, 1-1 Naka-ku, Sakai, Osaka 599-8531, Japan}
\author{Jun-ichiro Kishine}
\affiliation{Institute for Molecular Science, National Institutes of Natural Sciences, Okazaki 444-8585, Japan}
\affiliation{Division of Natural and Environmental Sciences, The Open University of Japan, Chiba 261-8586, Japan}
\date{\today}

\begin{abstract}
  In chiral crystals, two types of phonon angular momenta have been introduced.
  One is crystal angular momentum (CAM) arising from the rotational or screw-rotational symmetry and the other is mechanical angular momentum (MAM) associated with the circular motion of atomic displacements about equilibrium positions.
  Recently, the electron-phonon coupling that respects the screw-rotational symmetry is derived, whereby the CAM between electrons and phonons is interconverted.
  Here, we show that, in addition to CAM, MAM can also be converted to the electronic degrees of freedom by deriving a second-order perturbative Hamiltonian proportional to phonon MAM.
  This finding highlights that the electronic motion is directly affected by phonon MAM, and consequently, that phonon degrees of freedom can play a crucial role in phenomena related to electronic orbital polarization.
\end{abstract}
\maketitle

\section{Introduction}~\label{sec:intro}

Chirality is defined as a lack of improper rotational symmetry.
In chiral crystals, this symmetry breaking induces a phonon-band splitting between modes of opposite crystal angular momentum (CAM)~\cite{zhang2015Chiral,zhang2022Chiral,tsunetsugu2023Theory,kato2023Note}, which was recently confirmed by Raman and resonant inelastic x-ray scatterings using the circularly polarized light~\cite{ishito2022Truly,ishito2023Chiral,oishi2024Selective,ueda2023Chiral}.
CAM, also referred to as pseudo angular momentum, is a conserved quantity arising from the discrete rotational or screw-rotational symmetry of the crystal, which was originally introduced for the classification of electronic bands~\cite{bozovic1984Possible}.
CAM is distinct from mechanical angular momentum (MAM)~\cite{zhang2014Angular}, which is associated with circular motion of atomic displacements about equilibrium positions and thus with phonon spin angular momentum~\cite{vonsovskii1962Phonon}, in analogy to the photon spin angular momentum of circularly polarized light.
The chirality-induced phonon-band splitting is expected to provide a route for the interconversion of angular momentum between electrons and phonons via electron-phonon interactions.
This consideration may be relevant to clarifying the role of phonon degrees of freedom in chirality-induced spin selectivity~\cite{bloom2024Chiral,fransson2020Vibrational,du2020Vibrationenhanced,fransson2023Chiral,funato2024ChiralityInduced,inui2020ChiralityInduced} and other chirality-induced phenomena~\cite{rikken2002Observation,furukawa2017Observation,ohe2024ChiralityInduced,bousquet2025Structural}.

In this study, we focus on the phonon MAM and its interaction with electronic degrees of freedom, while neglecting electronic spin, through the investigation of the electron-phonon interaction inherent in chiral crystals. Recently, characteristic features of the microscopic electron-phonon interaction in chiral systems have been derived by two of the present authors~\cite{tateishi2025Electron,tateishi2026Microscopic}: Screw-rotational symmetry is respected, whereby the appropriate rotational phase factors of equivalent atoms in the unit cell~\cite{hu2024Electronic} are preserved. As a consequence, selection rules that describe the interconversion of CAM between electrons and phonons are derived. Additionally, in contrast to conventional theoretical frameworks~\cite{rossler2009Solid}, transverse phonons, thereby mechanically circulating phonons, couple to electrons. This latter point motivates us to explore the possibility of interconversion between phonon MAM and electronic degrees of freedom.

The present work develops this previous study to elucidate the effect of the phonon MAM on an electronic system.
For this purpose, second-order perturbation theory is employed via the Schrieffer-Wolff transformation~\cite{schrieffer1966Relation,winkler2003Spin}, whereby direct coupling of electrons to the phonon MAM is demonstrated.
This result establishes that not only the phonon CAM but also the phonon MAM can be converted to an electronic system and thereby advances our understanding of various electronic and phononic phenomena observed in chiral systems.

The remaining part of this paper is organized as follows:
In Sec.~\ref{sec:model}, we introduce a model single helix and its Hamiltonian consisting of the electronic and phononic degrees of freedom.
In Sec.~\ref{sec:SW}, we perform the Schrieffer-Wolff transformation to obtain the second-order perturbation terms of the electron-phonon Hamiltonian, and clarify that a second-order term is proportional to the phonon MAM.
We also discuss the limitation and generality of our result and briefly mention its effect on the electronic orbital angular momentum.
Section~\ref{sec:conclude} is devoted to the concluding remarks.

\section{Model}~\label{sec:model}

For simplicity, we consider a single helix in chiral tellurium (\ce{Te}) belonging to the space group $P3_121$.
The helix is assumed to consist of $N$ unit cells, and atoms along the helix are indexed by an integer $\ell$ ($=1,2,\ldots,3N$).
The equilibrium position of the $\ell$th atom is given by $\bm{R}_\ell = ( \rho\cos((\ell-1)\alpha), \rho\sin((\ell-1)\alpha), c(\ell-1)/3 )^\top$,
where $\rho$ is the helix radius, $c$ is the lattice constant along the $z$ axis, and $\alpha = 2\pi/3$ is the rotation angle.
The helix is taken to belong to the line group $L3_1 = \{\hat{\mathcal{R}}^p \mid p = 0, \ldots, 3N-1\}$~\cite{bozovic1984Possible,bozovic1978Irreducible}, generated by the threefold screw operator $\hat{\mathcal{R}} = (\hat{C}_3 \mid c/3)$, which is expressed as a combination of the threefold rotation $\hat{C}_3$ and the $c/3$ translation along the $z$ axis.
The irreducible representation (irrep.) of this group is specified by the crystal wavenumber $k = 2\pi n/cN$ with $n = 0, 1, \ldots, N-1$ and by the CAM $m = 0, \pm 1$.
Hereafter, these are collectively denoted by $\Gamma = (k,m) = (k_\Gamma,m_\Gamma)$.
The character of the irrep.~$\Gamma$ for $\hat{\mathcal{R}}^p$ is given by $\chi_\Gamma(p) = e^{-ip(kc/3+m\alpha)}$.
Reflection about the $xz$ plane transforms the helix to its enantiomer belonging to the line group $L3_2$, which corresponds to inverting the rotation angle, $\alpha =2\pi/3 \to -2\pi/3$.
This operation also inverts CAM, $m \to -m$, thereby changing $\Gamma$ to $\bar{\Gamma}=(k,-m)$. These pair states are particularly important for identifying the chirality-induced band splittings~\cite{zhang2015Chiral,zhang2022Chiral,tsunetsugu2023Theory,kato2023Note}.
The $L3_2$ counterpart of subsequent results of $L3_1$ is easily obtained by changing the irreps.~$\Gamma = (k,m)$ to the irreps.~$\bar{\Gamma}=(k,-m)$.
In accordance with the electronic case, the phononic system is subject to the crystal symmetry, and its states are thus expressed by the irrep. of the group.

The Hamiltonian of the electron-phonon coupled system is decomposed into three,
$\hat{H} = \hat{H}_\mathrm{el} + \hat{H}_\mathrm{ph} + \hat{H}_\mathrm{el-ph}$.
First, the electronic Hamiltonian is written in diagonal form as
\begin{equation}
  \hat{H}_\mathrm{el}
  = \sum_{\Gamma,\mu} E_{\Gamma\mu} \ketbra*{\Psi_{\Gamma}^\mu}{\Psi_{\Gamma}^\mu},
\end{equation}
where $E_{\Gamma\mu}$ and $\ket*{\Psi_{\Gamma}^\mu}$ denote the band energy and the corresponding eigenstate of the $\mu$th branch in the irrep.~$\Gamma$, respectively (see Appendix~\ref{subsec:Hel}).
In Fig.~\ref{fig:band_el}, the possible band structure is depicted, where the system is assumed to be composed of the $p$-atomic orbitals including only the nearest-neighbor coupling. The bands are degenerate for $m=+1$ and $-1$ at $k=0$ and for $m=-1$ and $0$ for $k=\pi/c$ correctly reflecting the symmetry of the line group~\cite{bozovic1984Possible}. Additionally, reflection about $k=0$ transforms the bands with $m=+1$ onto $m=-1$ (and vice versa), and with $m=0$ onto itself as a consequence of the time-reversal symmetry.

\begin{figure}
  \centering
  \includegraphics[width=81mm]{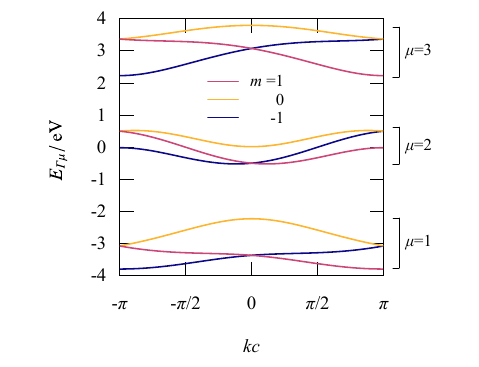}
  \caption{Electronic band energy, $E_{\Gamma\mu}$, of $p$-orbitals system in $L3_1$. The branch index $\mu$ is labeled in ascending order. In numerical calculation, we set $\rho = \qty{0.12}{\nm}$ and $c = \qty{0.591}{\nm}$,
    thereby the Slater-Koster overlap integrals are estimated to be $V_{pp\sigma} = \qty{3.02}{\eV}$ and $V_{pp\pi} = \qty{-0.75}{\eV}$.}
  \label{fig:band_el}
\end{figure}

The phononic system is described by the displacement vector about equilibrium position with the irrep.~$\Gamma$, $\bm{u}_\Gamma = \hat{u}_\Gamma \bm{v}_\Gamma$, and its conjugate momentum operator, $\bm{p}_\Gamma = \hat{p}_\Gamma \bm{v}_{-\Gamma}$, where $\hat{u}_\Gamma$ and $\hat{p}_\Gamma$ satisfy the commutation relation $[\hat{u}_\Gamma, \hat{p}_{\Gamma'} ] = i\hbar \delta_{\Gamma,\Gamma'}$, and $\bm{v}_\Gamma$ is the eigenvector of the dynamical matrix with the eigenvalue $\omega_\Gamma^2$ being the square of the frequency (the phononic band).
Here, the time-reversal operation connects a state with $\Gamma$ to that with $-\Gamma = (-k,-m)$. Because the time-reversal symmetry is preserved, the eigenvector satisfies the relation $\bm{v}_\Gamma^\ast = \bm{v}_{-\Gamma}$, which is used later to extract the phonon MAM from the electron-phonon interaction.
The phononic Hamiltonian is then given by
\begin{equation}
  \hat{H}_\mathrm{ph}
  = \sum_\Gamma \ab( \frac{1}{2M} \hat{p}_{-\Gamma} \hat{p}_\Gamma
  + \frac{M\omega_\Gamma^2}{2} \hat{u}_{-\Gamma}\hat{u}_\Gamma ),
\end{equation}
where $M$ is the atomic mass.
In this study, only the phononic bands within the single branch is considered, and therefore, the branch index is omitted.

Finally, we introduce the electron--phonon coupling Hamiltonian, $\hat{H}_\mathrm{el-ph}$, as the first-order contribution from phononic displacement of the electronic intersite couplings.
This Hamiltonian is derived in Ref.~\cite{tateishi2025Electron} (for details of the derivation, see Appendix~\ref{subsec:Hep}),
and is written as
\begin{equation}
  \hat{H}_\mathrm{el-ph}
  = \sum_\Gamma \hat{u}_{-\Gamma} \hat{W}_\Gamma,
\end{equation}
where
\begin{equation}
  \hat{W}_\Gamma
  = \sum_{\gamma,\mu\mu'} V_{\gamma \mu,\gamma+\Gamma \mu'} \ketbra*{\Psi_\gamma^\mu}{\Psi_{\gamma+\Gamma}^{\mu'}}.
\end{equation}
Here, we stress that the coupling constant $V_{\gamma \mu,\gamma+\Gamma \mu'}$ is a projection of the coupling vector $\bm{V}_{\gamma \mu,\gamma+\Gamma \mu'}$ onto the phonon polarization vector $\bm{v}_\Gamma$ as
\begin{equation}
  V_{\gamma \mu,\gamma+\Gamma \mu'}
  = \bm{v}_\Gamma \cdot \bm{V}_{\gamma \mu,\gamma+\Gamma \mu'}.
  \label{eq:el_coup_const}
\end{equation}
Thus, not only the longitudinal but also the transverse components of $\bm{v}_\Gamma$ enter the coupling, indicating that the circulating phonons carrying angular momentum can influence electronic motion.
This feature arises because the Hamiltonian is derived while respecting the screw-rotational symmetry of the helix, whose action induces both longitudinal and transverse motions,
whereas the conventional framework relies on lattice translational symmetry and involves only the longitudinal motion.
Moreover, the coupling $\bm{v}_\Gamma \cdot \bm{V}_{\gamma \mu,\gamma+\Gamma \mu'}$ directly reflects the selection rule:
electronic transition from the irrep.~$\gamma+\Gamma$ to $\gamma$ are mediated only by the phonon modes of the irrep.~$\Gamma$.
The form in Eq.~\eqref{eq:el_coup_const} is necessary to extract the phonon MAM from the second-order perturbative Hamiltonian.
All Hamiltonians introduced above satisfy time-reversal symmetry, as expressed by the relations
$E_{\Gamma\mu} = E_{-\Gamma\mu}$, $\omega_\Gamma = \omega_{-\Gamma}$, and $\bm{V}_{\gamma\mu,\gamma'\mu'} = \bm{V}_{-\gamma'\mu',-\gamma\mu}$.
Furthermore, chiral symmetry breaking is reflected in the inequalities
$E_{\Gamma\mu} \ne E_{\bar{\Gamma}\mu}$, $\omega_{\Gamma} \ne \omega_{\bar{\Gamma}}$,
and $\bm{V}_{\gamma\mu,\gamma'\mu'} \ne \bm{V}_{\bar{\gamma}\mu,\bar{\gamma'}\mu'}$.

\section{Schrieffer-Wolff Transformation}~\label{sec:SW}

To elucidate phonon effects on electronic dynamics in the electron-phonon coupled system,
the following unitary (Schrieffer-Wolff) transformation~\cite{schrieffer1966Relation,winkler2003Spin} is performed:
\begin{align}
  \hat{H}
  \longrightarrow \tilde{H}
   & = e^{\hat{S}} \hat{H} e^{-\hat{S}}
  \notag                                \\
   & = \hat{H} + \ab[ \hat{S}, \hat{H}]
  + \frac{1}{2}\ab[\hat{S}, \ab[ \hat{S}, \hat{H}] ]
  + \cdots,
\end{align}
where the operator $\hat{S}$ is assumed to take the form
\begin{equation}
  \hat{S}
  = \frac{i}{\hbar} \sum_\Gamma
  \ab( \hat{p}_{-\Gamma} \hat{F}_\Gamma
  + \hat{u}_{-\Gamma} \hat{G}_\Gamma ).
\end{equation}
To determine the electronic operators $\hat{F}_\Gamma$ and $\hat{G}_\Gamma$,
the condition
\begin{equation}
  \hat{H}_{\mathrm{el-ph}}
  + \ab[ \hat{S}, \hat{H}_{\mathrm{el}} + \hat{H}_{\mathrm{ph}} ]
  = 0.
  \label{eq:condition_SW}
\end{equation}
is imposed.
By neglecting higher-order contributions, the transformed Hamiltonian $\tilde{H}$ can be approximated to the second order in the electron-phonon coupling as
\begin{equation}
  \tilde{H}
  \approx \hat{H}_{\mathrm{el}} + \hat{H}_{\mathrm{ph}}
  + \hat{H}_2.
\end{equation}
where the second-order term is given by $\hat{H}_2 = [ \hat{S}, \hat{H}_{\mathrm{el-ph}} ]/2$.
Eq.~\eqref{eq:condition_SW} is satisfied when
\begin{equation}
  \left\{
  \begin{aligned}
     & \frac{i}{\hbar} \ab[ \hat{F}_\Gamma, \hat{H}_{\mathrm{el}}]
    - \frac{1}{M}\hat{G}_{-\Gamma} = 0
    \\
     & \frac{i}{\hbar} \ab[ \hat{G}_\Gamma, \hat{H}_{\mathrm{el}}]
    + \hat{W}_\Gamma
    + M\omega_\Gamma^2 \hat{F}_{-\Gamma} = 0
  \end{aligned}
  \right.
  \label{eq_for_FGamma_GGamma}
\end{equation}
holds.
By solving these equations,
the matrix elements of $\hat{F}_\Gamma$ and $\hat{G}_\Gamma$ are obtained as
\begin{align}
  (\hat{F}_\Gamma)_{\gamma\mu,\gamma'\mu'}
   & = g_{\gamma\mu,\gamma'\mu'}(\Gamma) (\hat{W}_{-\Gamma})_{\gamma\mu,\gamma'\mu'},
  \\
  (\hat{G}_\Gamma)_{\gamma\mu,\gamma'\mu'}
   & = -\frac{iM}{\hbar} E_{\gamma\mu,\gamma'\mu'}
  g_{\gamma\mu,\gamma'\mu'}(\Gamma) (\hat{W}_{\Gamma})_{\gamma\mu,\gamma'\mu'},
\end{align}
where $(\hat{A})_{\gamma\mu,\gamma'\mu'} = \braket*[3]{\Psi_\gamma^\mu}{\hat{A}}{\Psi_{\gamma'}^{\mu'}}$ denotes the matrix element of an electronic operator $\hat{A}$.
Here, $g_{\gamma\mu,\gamma'\mu'}(\Gamma)$ is defined as
\begin{equation}
  g_{\gamma\mu,\gamma'\mu'}(\Gamma)
  \equiv \frac{\hbar^2}{M} \frac{1}{(E_{\gamma\mu,\gamma'\mu'})^2 - (\hbar\omega_\Gamma)^2},
  \label{eq:g_factor}
\end{equation}
and $E_{\gamma\mu,\gamma'\mu'} = E_{\gamma\mu} - E_{\gamma'\mu'}$ is the band energy difference.

As a result, the second-order term is given as
\begin{equation}
  \hat{H}_2
  = \sum_\Gamma \hat{X}_{\Gamma}
  + \frac{i}{\hbar} \sum_{\Gamma\Gamma'} \ab(
  \hat{Y}_{\Gamma,\Gamma'} \hat{u}_{-\Gamma'} \hat{p}_{-\Gamma}
  + \hat{Z}_{\Gamma,\Gamma'} \hat{u}_{-\Gamma'} \hat{u}_{-\Gamma}
  )
  \label{eq:S_Helph_commute}
\end{equation}
with $\hat{X}_\Gamma = \hat{F}_\Gamma \hat{W}_{\Gamma}$,
$\hat{Y}_{\Gamma,\Gamma'} = [ \hat{F}_\Gamma, \hat{W}_{\Gamma'} ]$,
and $\hat{Z}_{\Gamma,\Gamma'} = [ \hat{G}_\Gamma, \hat{W}_{\Gamma'} ]$.

\subsection{Emergence of phonon MAM in electron-phonon coupling}

In the regime where the atomic mass is sufficiently larger than the electronic mass,
the electronic back-action on phonons is negligible.
This neglect is consistent with our second-order truncation: the back-action amounts to an electron-induced phonon self-energy of order $V_{\gamma\mu,\gamma'\mu'}^{2}$, which would contribute only at $\mathcal{O}(V_{\gamma\mu,\gamma'\mu'}^{4})$ because the effective Hamiltonian $\hat{H}_{2}$ is already the order of $\mathcal{O}(V_{\gamma\mu,\gamma'\mu'}^{2})$. The replacement of the phonon operators by their equilibrium expectation values is therefore required by consistency at second order, with the mass ratio providing the complementary adiabatic separation of timescales. We emphasize that no vertex correction is involved in the present scheme.
Hence, the phonon operators in Eq.~\eqref{eq:S_Helph_commute} can be safely replaced with their thermal equilibrium expectation values:
$\hat{u}_{-\Gamma'} \hat{p}_{-\Gamma} \to \aab*{\hat{u}_{-\Gamma'} \hat{p}_{-\Gamma}}$
and $\hat{u}_{-\Gamma'} \hat{u}_{-\Gamma} \to \aab*{\hat{u}_{-\Gamma'} \hat{u}_{-\Gamma}}$.
Namely, the phonon effects are treated at the mean-field level.
To evaluate these quantities,
the phonon operators are expressed in terms of the annihilation ($\hat{a}_\Gamma$) and the creation ($\hat{a}_\Gamma^\dagger$) operators, which satisfy $[\hat{a}_\Gamma, \hat{a}_{\Gamma'}^\dagger ] = \delta_{\Gamma,\Gamma'}$
as $\hat{u}_\Gamma = \sqrt{\hbar/2M\omega_\Gamma}(\hat{a}_\Gamma+\hat{a}_{-\Gamma}^\dagger)$ and $\hat{p}_\Gamma = i\sqrt{\hbar\omega_\Gamma M/2} (\hat{a}_\Gamma^\dagger - \hat{a}_{-\Gamma})$.
Using the relation $\aab*{\hat{a}_{\Gamma'}^\dagger \hat{a}_\Gamma} = \delta_{\Gamma\Gamma'} f(\omega_{\Gamma})$
for the Bose-Einstein distribution $f$,
Eq.~\eqref{eq:S_Helph_commute} becomes
\begin{align}
  \ab( \hat{H}_2 )_{\gamma\mu,\gamma'\mu'}
   & = \delta_{\gamma,\gamma'} \sum_{\Gamma} \left[
    (\hat{X}_\Gamma)_{\gamma\mu,\gamma\mu'} \right.
  \notag                                                                                                               \\
   & \left. + \ab( \hat{Y}_{\Gamma,\Gamma} + \frac{i}{M\omega_\Gamma} \hat{Z}_{\Gamma,-\Gamma})_{\gamma\mu,\gamma\mu'}
    \ab(f(\omega_\Gamma)+\frac{1}{2} ) \right].
  \label{eq:H2_classical}
\end{align}
Thus, only the diagonal elements with $\gamma = \gamma'$ contribute to the Hamiltonian,
which are given by
\begin{align}
   & (\hat{X}_\Gamma)_{\gamma\mu,\gamma\mu'}
  = \sum_{\mu''} g_{\gamma\mu,\gamma-\Gamma\mu''}(\Gamma)
  V_{\gamma\mu,\gamma-\Gamma\mu''} V_{\gamma-\Gamma\mu'',\gamma\mu'},
  \\
   & (\hat{Z}_{\Gamma,-\Gamma})_{\gamma\mu,\gamma\mu'}
  \notag                                               \\
   & = -\frac{iM}{\hbar} \sum_{\mu''} \left[
    E_{\gamma\mu,\gamma+\Gamma\mu''} g_{\gamma\mu,\gamma+\Gamma\mu''}(\Gamma)
    V_{\gamma\mu,\gamma+\Gamma\mu''} V_{\gamma+\Gamma\mu'',\gamma\mu'} \right.
  \notag                                               \\
   & \quad + \left.
    E_{\gamma\mu',\gamma-\Gamma\mu''} g_{\gamma\mu',\gamma-\Gamma\mu''}(\Gamma)
    V_{\gamma\mu,\gamma-\Gamma\mu''} V_{\gamma-\Gamma\mu'',\gamma\mu'} \right],
\end{align}
and $(\hat{Y}_{\Gamma,\Gamma})_{\gamma\mu,\gamma\mu'} = (\hat{X}_\Gamma)_{\gamma\mu,\gamma\mu'} - (\hat{X}_{-\Gamma})_{\gamma\mu',\gamma\mu}^\ast$.
Hence, all these elements include the factor $V_{\gamma\mu,\gamma\pm\Gamma\mu''}V_{\gamma\pm\Gamma\mu'',\gamma\mu'}=(\bm{v}_{\pm\Gamma}\cdot\bm{V}_{\gamma\mu,\gamma\pm\Gamma\mu''})
  (\bm{v}_{\mp\Gamma}\cdot\bm{V}_{\gamma\pm\Gamma\mu'',\gamma\mu'})$.
Using the vector identity
\begin{align}
   & (\bm{v}\cdot\bm{V})(\bm{v}'\cdot\bm{V}')
  \notag                                                                       \\
   & = \sum_{a,b} V_a v_a^\ast (v_b')^\ast V_b'
  \notag                                                                       \\
   & = \sum_{a,b} V_a \ab[ \frac{v_a^\ast (v_b')^\ast+v_b^\ast (v_a')^\ast}{2}
    + \frac{v_a^\ast (v_b')^\ast-v_b^\ast (v_a')^\ast}{2}
  ] V_b',
\end{align}
we can separate the factor into the symmetric and antisymmetric parts of the phonon polarization tensor $\bm{v}_{-\Gamma}\bm{v}_\Gamma^\top$ as
\begin{align}
  V_{\gamma\mu,\gamma\pm\Gamma\mu''}V_{\gamma\pm\Gamma\mu'',\gamma\mu'}
   & = \frac{1}{2}\bm{V}_{\gamma\mu,\gamma\pm\Gamma\mu''}^\top
  \mathcal{P}_{\pm\Gamma} \bm{V}_{\gamma\pm\Gamma\mu'',\gamma\mu'}
  \notag                                                                                                                                       \\
   & \quad \pm \frac{i}{2\hbar} \bm{L}_\Gamma \cdot (\bm{V}_{\gamma\mu,\gamma\pm\Gamma\mu''} \times \bm{V}_{\gamma\pm\Gamma\mu'',\gamma\mu'}),
  \label{eq:V_decomp}
\end{align}where
$\mathcal{P}_\Gamma = \bm{v}_{-\Gamma} \bm{v}_{\Gamma}^\top + \bm{v}_{\Gamma} \bm{v}_{-\Gamma}^\top$ is the symmetric phonon polarization tensor and
the time-reversal symmetric relation $\bm{v}_\Gamma = \bm{v}_{-\Gamma}^\ast$ is used.
Here, the antisymmetric part is rewritten using the phonon MAM
\begin{equation}
  \bm{L}_\Gamma
  = -i\hbar \bm{v}_{-\Gamma} \times \bm{v}_\Gamma,
  \label{eq:def_phonon_MAM}
\end{equation}
in the irrep.~$\Gamma$ (for details, see Appendix~\ref{subsec:Lph}).
This result establishes that the electron-phonon coupling contains a term directly proportional to the phonon MAM.
Recently, the term ``axial phonon'' has been introduced for a phonon mode carrying CAM and/or MAM~\cite{juraschek2025Chiral}. Thus, Eqs.~\eqref{eq:H2_classical}-\eqref{eq:def_phonon_MAM} can be regarded as the direct coupling between electrons and axial phonons. Because phonon modes with nonzero wavenumber carrying angular momentum is clearly chiral, these also include a direct coupling to chiral axial phonons.
It should be noted that this coupling directly reflects the presence of the transverse components of phonon displacement in the electron-phonon interaction, $\hat{H}_\mathrm{el-ph}$.
Otherwise, only the $z$-component of $\bm{V}_{\gamma\mu,\gamma\pm\Gamma\mu'}$ is involved in the quantity $\bm{V}_{\gamma\mu,\gamma\pm\Gamma\mu''} \times \bm{V}_{\gamma\pm\Gamma\mu'',\gamma\mu'}$, which is identically equal to zero.
For the line group $L3_2$, $\bm{v}_\Gamma$ and $\bm{V}_{\gamma\mu,\gamma'\mu'}$ in Eq.~\eqref{eq:V_decomp} are replaced by $\bm{v}_{\bar{\Gamma}}$ and $\bm{V}_{\bar{\gamma}\mu,\bar{\gamma'}\mu'}$, respectively, while the MAM operator $\bm{L}_\Gamma$ is replaced by $-\bm{L}_{\bar{\Gamma}}$.

\subsection{Band renormalization}~\label{sec:band_el}

\begin{figure}[t]
  \centering
  \includegraphics[width=81mm]{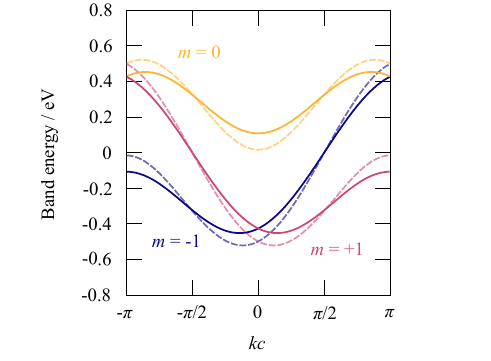}
  \caption{Electronic energy bands without (dashed lines) and with (solid lines) the electron-phonon interaction, respectively.
    The Parameters are set to $T = \qty{300}{\kelvin}$ and $\hbar^2\zeta^2/M = \qty{3.0d-4}{\eV}$, which corresponds to $\zeta = \qty{1d10}{\meter^{-1}}$.}
  \label{fig:band}
\end{figure}

The Schrieffer-Wolff transformation can be regarded as a form of quasidegenerate perturbation theory~\cite{winkler2003Spin}, in which the low-energy electronic subspace is retained while the remaining states are perturbatively integrated out to construct an effective Hamiltonian. For the $p$-orbital case shown in Fig.~\ref{fig:band_el}, $\hat{H}_2$ is restricted to the electronic bands with branch index
$\mu=2$, which lie near the Fermi energy, $E_\mathrm{F}=0$, and are energetically well separated from the bands with $\mu \neq 2$. Accordingly, the second-order effective Hamiltonian $\hat{H}_2$ contains only contributions arising from virtual interbranch
transitions mediated by the electron-phonon interaction. This restriction prevents the resonant enhancement associated with the energy denominator in Eq.~\eqref{eq:g_factor}, because the energy separation between different electronic branches is larger than the
relevant phonon energy scale. Consequently, the corresponding energy denominators remain finite and cannot approach resonance through phonon excitation, ensuring the validity of the second-order perturbative treatment. Intrabranch transitions first contribute at third order in perturbation theory.

In this case, the second-order effective Hamiltonian takes the form
\begin{align}
  (\hat{H}_2)_{\gamma\mu,\gamma\mu}
   & = \sum_\Gamma \sum_{\mu' \ne \mu}
  g_{\gamma\mu,\gamma+\Gamma\mu'}(\Gamma) \vab*{V_{\gamma\mu,\gamma+\Gamma\mu'}}^2
  \notag                               \\
   & \quad \times
  \ab[1 + \frac{2 E_{\gamma\mu,\gamma+\Gamma\mu'}}{\hbar\omega_\Gamma}
    \ab( f(\omega_\Gamma)+\frac{1}{2} ) ].
  \label{eq:H2_band}
\end{align}

Figure~\ref{fig:band} depicts the electronic band dispersions without (dashed lines) and with (solid lines) coupling to the lowest optical modes, respectively. Although the electron-phonon coupling induces appreciable changes in the band energies, suggesting a non-negligible contribution from the phonon MAM, the overall band structure remains largely unchanged. This is because, as shown in Fig.~\ref{fig:band_el}, the interbranch electronic energy separation is much larger than the phononic energy scale.
We also note that, in Eq.~\eqref{eq:H2_band}, the condition $E_{\gamma\mu,\gamma+\Gamma\mu'}/\hbar\omega_\Gamma \gg 1$ in the second term implies that the temperature dependent term dominates $\hat{H}_2$. Furthermore, the coupling strength is inversely proportional to the atomic mass, $M$. Therefore, the phonon MAM-induced effects are expected to be enhanced in high-temperature regime and in crystals composed of light elements.

The limitation of the perturbative treatment does not affect the general structure of the coupling to the phonon MAM.
More generally, in the higher-order perturbation theory,
conservation of the phonon occupation number requires $\bm{v}_\Gamma$ to appear together with its time-reversal partner, $\bm{v}_{-\Gamma}$.
Their contribution can be combined into the factor $\bm{v}_{-\Gamma} \times \bm{v}_\Gamma \propto \bm{L}_\Gamma$.
Consequently, irrespective of the order of perturbative theory,
the electronic degrees of freedom couple directly to the phonon MAM in chiral crystals.

\subsection{Effect on electronic orbital angular momentum}

Finally in this section, we briefly discuss a possible influence of the electron-phonon coupling on the electronic OAM~\cite{xiao2010Berry}.
The OAM operator is defined by
\begin{equation}
  \hat{\bm{L}}_\mathrm{el}
  = \frac{m_\mathrm{e}}{2}
  \ab( \hat{\bm{r}} \times \hat{\bm{v}} - \hat{\bm{v}} \times \hat{\bm{r}} ),
\end{equation}
where $m_\mathrm{e}$ is the mass,
$\hat{\bm{r}}$ is the position operator, and
$\hat{\bm{v}} = (i/\hbar)[\hat{H}_\mathrm{el},\hat{\bm{r}}]$ is the velocity operator of an electron.
For our purpose, it is sufficient to discuss the $z$ component of OAM although we can formulate the other components using a similar procedure~\cite{xiao2010Berry,pezo2022Orbital}.
Let $\ket*{\phi_\ell^\lambda}$ denote the atomic orbital $\lambda$ centered at $\bm{R}_\ell$.
Assuming that $\ket*{\phi_\ell^\lambda}$ is spatially localized, and hence, is the eigenstate of the position operator as $\hat{\bm{r}} \ket*{\phi_\ell^\lambda} = \bm{R}_\ell \ket*{\phi_\ell^\lambda}$, the matrix element of the $z$ component of OAM becomes (see Appendix~\ref{subsec:Lel} for the derivation)
\begin{equation}
  \braket*[3]{\Psi_\Gamma^\mu}{\hat{L}_{\mathrm{el},z}}{\Psi_\Gamma^\mu}
  = -m_\mathrm{e} \sum_{\ell,\lambda,\lambda'}
  (\bm{R}_{1+\ell} \times \bm{R}_1 )_z J_{\ell;\Gamma\mu}^{\lambda'\lambda},
  \label{eq:Lz_el}
\end{equation}
where $(\bm{R}_{1+\ell} \times \bm{R}_1 )_z = -\rho^2\sin(\ell\alpha)$ is the area spanned by the vectors associated with the hopping from site $1$ to $1+\ell$, and
\begin{equation}
  J_{\ell;\Gamma\mu}^{\lambda'\lambda}
  = -\frac{i}{\hbar} (U_\Gamma)_{\lambda'\mu}^\ast (U_\Gamma)_{\lambda\mu}
  e^{i\ell\Phi_\Gamma^{\lambda'}} \braket*[3]{\phi_{1+\ell}^{\lambda'}}{\hat{H}_\mathrm{el}}{\phi_1^\lambda},
  \label{eq:rate_Lz}
\end{equation}
is the corresponding hopping rate of $\ket*{\Psi_\Gamma^\mu}$ from orbital $\lambda$ to $\lambda'$.
Therefore, OAM can be interpreted as the sum of the hopping rate weighed by the hopping area.
In general, OAM for Bloch electrons is divided into intra-atomic and interatomic contributions~\cite{pezo2022Orbital,burgosatencia2024Orbital}.
Whereas the former is interpreted as the average magnetic quantum number of the eigenstate described by the coupling matrix $U_\Gamma$, the latter is associated with the circular trajectory of the electronic motion, which is described by the transition matrix elements $\braket*[3]{\phi_{1+\ell}^{\lambda'}}{\hat{H}_\mathrm{el}}{\phi_1^\lambda}$.
When we include the contribution from electron-phonon interaction, $\hat{H}_\mathrm{el}$ is replaced by $\hat{H}_\mathrm{el}+\hat{H}_2$, which indicates that the electron-phonon interaction, and hence the phononic MAM, directly affect the OAM through modifying the electronic couplings.
Because electron transport via next-nearest-neighbor coupling has an interatomic contribution opposite to that via nearest-neighbor coupling, the electron-phonon interaction may either increase or decrease the OAM, as determined by a more detailed analysis.

Equations~\eqref{eq:Lz_el} and \eqref{eq:rate_Lz} also suggest an experimental route to detect the direct electronic coupling to phonon MAM in our formulation: this coupling should appear as a modulation of the electronic OAM in response to phonon excitation and the associated modulation of phonon MAM, for example by circularly polarized terahertz waves~\cite{luo2023Large,basini2024Terahertz,davies2024Phononic}.

\section{Concluding Remarks}~\label{sec:conclude}

In this study, we analyzed the electron-phonon coupling in chiral crystals.
As a result of the screw-rotational symmetry, both the longitudinal and transverse modes of phonons are involved in the coupling, and hence, the phononic MAM is coupled directly to the electronic systems.
This indicates that in chiral systems the angular momentum of the electronic and phononic degrees of freedom can be interconverted, which has an implication of the phononic enhancement of the orbital polarization~\cite{liu2021Chiralitydriven,cho2026ChiralityInduced,sato2026Orbital,yao2026Dynamical,nabei2026Orbital}.
Furthermore, taking account of spin-orbit coupling, the present scheme of orbital polarization may be relevant to experimentally suggested angular momentum conversion from phonon to spin~\cite{ohe2024ChiralityInduced} or phonon-assisted spin Seebeck~\cite{kim2023Chiralphononactivated} responses.
The phononic enhancement of orbital polarization can be verified by detecting the electronic angular momentum induced by the selective excitation of phonons with circularly polarized terahertz waves.
Extension of our result to the general chiral crystals and molecules and exploration of the possibility of the electronic control of the phonon MAM will be left for future studies.

\begin{acknowledgments}
  We thank Tomomi Tateishi for fruitful discussions.
  This work was supported by the JSPS KAKENHI Grants No.~21H05019, No.~22K04863, No.~22H05132, No.~23H00091, No.~25K00962, No.~25H02149, and No.~26H02234,
  by JSPS International Joint Research Program 300 JRP-LEAD with UKRI (Grant No.~JPJSJRP20241710),
  by the OML Project grant by the National Institutes of Natural Sciences (NINS Program No.~OML012301),
  and by JST ERATO Grant No.~JPMJER2503, Japan.
\end{acknowledgments}

\appendix
\section{Microscopic Description of the Model}~\label{sec:Hep_derivation}
\subsection{Electronic Hamiltonian}~\label{subsec:Hel}

Let $\ket*{\phi_\ell^\lambda}$ denote an electronic state with atomic orbital $\lambda$ centered at $\bm{R}_\ell$.
This state can be decomposed into a symmetry-adapted basis $\ket*{\phi_\Gamma^\lambda}$,
which transforms as $\hat{\mathcal{R}} \ket*{\phi_\Gamma^\lambda} = e^{i(kc/3+m\alpha)}\ket*{\phi_\Gamma^\lambda}$,
according to
\begin{equation}
  \ket*{\phi_\ell^\lambda}
  = \frac{1}{\sqrt{3N}} \sum_\Gamma
  e^{i(\ell-1)\Phi_\Gamma^\lambda} \ket*{\phi_\Gamma^\lambda},
  \label{eq:c_el_site_to_IR}
\end{equation}
where $\Phi_\Gamma^\lambda \coloneq kc/3 + (m+m_\lambda)\alpha$ is a phase factor
and $m_\lambda$ is the magnetic quantum number of orbital $\lambda$.
The electronic Hamiltonian is then expressed as
\begin{align}
  \hat{H}_\mathrm{el}
   & = \sum_{\ell\ell'\lambda\nu} t_{\ell',\ell}^{\lambda\nu}
  \ketbra*{\phi_{\ell'}^\lambda}{\phi_\ell^\nu}
  \notag                                                            \\
   & = \frac{1}{3N} \sum_{\ell\ell'\lambda\nu} \sum_{\Gamma\Gamma'}
  t_{1+\ell'-\ell,1}^{\lambda\nu} e^{i(\ell'-\ell)\Phi_{\Gamma'}^\lambda}
  \notag                                                            \\
   & \quad \times \chi_{\Gamma'}^\ast(\ell-1) \chi_\Gamma(\ell-1)
  \ketbra*{\phi_{\Gamma'}^\lambda}{\phi_{\Gamma}^\nu}
  \notag                                                            \\
   & = \sum_{\lambda\nu\Gamma} t_\Gamma^{\lambda\nu}
  \ketbra*{\phi_{\Gamma}^\lambda}{\phi_{\Gamma}^\nu},
  \label{eq:Hel_appnd}
\end{align}
where $t_\Gamma^{\lambda\nu} = \sum_\ell t_{1+\ell,1}^{\lambda\nu} e^{i\ell\Phi_\Gamma^\lambda}$
is the generalized Fourier transform of the electronic coupling $t_{\ell',\ell}^{\lambda\nu} = \braket*[3]{\phi_{\ell'}^\lambda}{\hat{H}_\mathrm{el}}{\phi_\ell^\nu}$.
To derive Eq.~\eqref{eq:Hel_appnd},
the screw symmetry relation $t_{\ell',\ell}^{\lambda\nu} = e^{-i(\ell-1)(m_\lambda-m_\nu)\alpha} t_{1+\ell'-\ell,1}^{\lambda\nu}$ and the great orthogonality theorem for characters are used.
By solving the eigenvalue equation in the orbital space,
$t_\Gamma U_{\Gamma,\mu} = E_{\Gamma\mu} U_{\Gamma,\mu}$,
and defining the unitary matrix as $U_\Gamma = (U_{\Gamma,\mu=1}, U_{\Gamma,\mu=2}, \ldots)$,
the eigenstates are obtained as $\ket*{\Psi_\Gamma^\mu} = \sum_\lambda (U_\Gamma)_{\lambda\mu} \ket*{\phi_\Gamma^\lambda}$, which yields
\begin{equation}
  \hat{H}_\mathrm{el}
  = \sum_{\Gamma\mu} E_{\Gamma\mu} \ketbra*{\Psi_\Gamma^\mu}{\Psi_\Gamma^\mu}.
\end{equation}

\subsection{Electron-phonon coupling Hamiltonian}~\label{subsec:Hep}

The atomic position of the $\ell$th atom, $\bm{r}_\ell$,
is written as the sum of the equilibrium position and the displacement $\bm{u}_\ell$, $\bm{r}_\ell = \bm{R}_\ell + \bm{u}_\ell$.
This displacement modulates the electronic couplings as
\begin{equation}
  t^{\lambda\nu}(\bm{r}_{\ell'}-\bm{r}_\ell)
  \approx t_{\ell',\ell}^{\lambda\nu}
  + (\bm{u}_{\ell'}-\bm{u}_\ell) \cdot
  \left. \diffp{t^{\lambda\nu}(\bm{r})}{\bm{r}} \right\rvert_{\bm{r}=\bm{R}_{\ell'}-\bm{R}_\ell},
  \label{eq:transfer_integral_phonon}
\end{equation}
where the first term, defined by $t_{\ell',\ell}^{\lambda\nu} \equiv t^{\lambda\nu}(\bm{R}_{\ell'}-\bm{R}_\ell)$,
and the second term contributes to the electronic and the electron-phonon coupling Hamiltonians, respectively.
Following Refs.~\cite{Friedel1970,tateishi2025Electron}, the spatial derivative of $t^{\lambda\nu}$ is treated as
$\difsp{t^{\lambda\nu}(\bm{r})}{\bm{r}}|_{\bm{r}=\bm{R}_{\ell'}-\bm{R}_\ell} = -\zeta t_{\ell',\ell}^{\lambda\nu} \bm{Q}_{\ell',\ell}$, where $\zeta$ characterizes the spatial extent of the electronic coupling and $\bm{Q}_{\ell',\ell} = (\bm{R}_{\ell'}-\bm{R}_\ell)/\vab*{\bm{R}_{\ell'}-\bm{R}_\ell}$ is the normalized bonding vector.

For chiral systems with screw-rotational symmetry, it is convenient to employ the circular basis defined by
$\bm{e}^{(\pm)} = (\bm{e}_x \mp i\bm{e}_y)/\sqrt{2}$ and $\bm{e}^{(0)} = \bm{e}_z$.
Each basis vector is the eigenvector of threefold rotational operator as $\hat{C}_3 \bm{e}^{(s)} = e^{im_s\alpha} \bm{e}^{(s)}$, where $m_s$ ($m_\pm = \pm1$, and $m_0 = 0$) is the helicity of the phonon spin angular momentum in analogy with the photon spin angular momentum of circularly polarized light.
Using the circular basis,
the bonding vector is expanded as $\bm{Q}_{\ell',\ell} = \sum_s q_{\ell'-\ell}^{(s)} e^{im_s \theta_{\ell'\ell}} \bm{e}^{(s)}$, where the coefficients are given by $q_\ell^{(\pm)} = \pm i \cos\gamma_\ell / \sqrt{2}$ and $q_\ell^{(0)} = \sin\gamma_\ell$.
Here, $\gamma_\ell = \arctan(c\ell/6\rho\sin(\ell\alpha/2))$ is the helical angle, and $\theta_{\ell'\ell} = (\ell'+\ell-2)\alpha/2$.

Next, the phonon displacement vector is expanded in the circular basis as $\bm{u}_\ell = \sum_s u_\ell^{(s)} \bm{e}^{(s)}$, with coefficients
\begin{equation}
  u_\ell^{(s)}
  = \frac{1}{\sqrt{3N}} \sum_\Gamma e^{-i(\ell-1)\Omega_\Gamma^s} \hat{u}_\Gamma v_{\Gamma}^{(s)},
\end{equation}
where $\Omega_\Gamma^s = kc/3 + (m-m_s)\alpha$ is the phase factor,
$\hat{u}_\Gamma = \sqrt{\hbar/2M\omega_\Gamma} (\hat{a}_\Gamma+\hat{a}_{-\Gamma}^\dagger)$ is the quantized displacement operator, and $v_\Gamma^{(s)}$ is the $s$th component of the eigenvector of the dynamical matrix.
The inner product $(\bm{u}_{\ell'}-\bm{u}_\ell)\cdot\bm{Q}_{\ell',\ell}$ is then evaluated as
\begin{align}
  (\bm{u}_{\ell'}-\bm{u}_\ell)\cdot\bm{Q}_{\ell',\ell}
   & = \frac{1}{\sqrt{3N}} \sum_{\Gamma,s}
  e^{i(\ell'-\ell)m_s\alpha/2} ( e^{i(\ell'-\ell)\Omega_\Gamma^s}-1 )
  \notag                                   \\
   & \quad \times
  \hat{u}_{-\Gamma} (v_\Gamma^{(s)})^\ast \chi_\Gamma(\ell-1) q_{\ell'-\ell}^{(s)}.
  \label{eq:u_dot_Q}
\end{align}

From Eq.~\eqref{eq:transfer_integral_phonon}, we can define the electron--phonon coupling Hamiltonian as
\begin{equation}
  \hat{H}_\mathrm{el-ph}
  = - \sum_{\ell\ell'\lambda\nu} \zeta t_{\ell',\ell}^{\lambda\nu}
  (\bm{u}_{\ell'}-\bm{u}_\ell) \cdot \bm{Q}_{\ell',\ell}
  \ketbra*{\phi_{\ell'}^\lambda}{\phi_\ell^\nu}.
\end{equation}
Substituting Eqs.~\eqref{eq:c_el_site_to_IR} and \eqref{eq:u_dot_Q} yields
\begin{align}
  \hat{H}_\mathrm{el-ph}
  = & -\frac{\zeta}{(3N)^{3/2}} \sum_{\ell\ell'\lambda\nu}
  \sum_{\Gamma,s} \sum_{\gamma,\gamma'}
  t_{\ell',\ell}^{\lambda\nu} q_{\ell'-\ell}^{(s)} \hat{u}_{-\Gamma} (v_\Gamma^{(s)})^\ast
  \notag                                                                          \\
    & \times e^{i(\ell'-\ell)m_s\alpha/2} ( e^{i(\ell'-\ell)\Omega_\Gamma^s}-1 )
  \chi_\Gamma(\ell-1)
  \notag                                                                          \\
    & \times e^{i(\ell'-1)\Phi_{\gamma}^\lambda} e^{-i(\ell-1)\Phi_{\gamma'}^\nu}
  \ketbra*{\phi_{\gamma}^\lambda}{\phi_{\gamma'}^\nu}
  \notag                                                                          \\
  = & -\frac{\zeta}{(3N)^{3/2}} \sum_{\ell\ell'\lambda\nu}
  \sum_{\Gamma,s} \sum_{\gamma,\gamma'}
  t_{1+\ell'-\ell,1}^{\lambda\nu} e^{i(\ell'-\ell)\Phi_{\gamma}^\lambda} q_{\ell'-\ell}^{(s)}
  \notag                                                                          \\
    & \times e^{i(\ell'-\ell)m_s\alpha/2} ( e^{i(\ell'-\ell)\Omega_\Gamma^s}-1 )
  \hat{u}_{-\Gamma} (v_\Gamma^{(s)})^\ast
  \notag                                                                          \\
    & \times \chi_\Gamma(\ell-1) \chi_\gamma(\ell-1) \chi_{\gamma'}^\ast(\ell-1)
  \ketbra*{\phi_\gamma^\lambda}{\phi_{\gamma'}^\nu}.
\end{align}
Changing the variables from $\ell$ and $\ell'$ to $\ell$ and $\ell''=\ell'-\ell$
and performing summation over $\ell$ with the great orthogonality theorem for characters,
we obtain
\begin{align}
  \hat{H}_\mathrm{el-ph}
  = & -\frac{\zeta}{\sqrt{3N}} \sum_{\ell''\lambda\nu}
  \sum_{\gamma,\Gamma,s}
  t_{1+\ell'',1}^{\lambda\nu} e^{i\ell''\Phi_{\gamma}^\lambda} q_{\ell''}^{(s)}
  \notag                                                              \\
    & \times e^{i\ell'' m_s\alpha/2} ( e^{i\ell''\Omega_\Gamma^s}-1 )
  \hat{u}_{-\Gamma} (v_\Gamma^{(s)})^\ast
  \ketbra*{\phi_\gamma^\lambda}{\phi_{\gamma+\Gamma}^\nu}.
\end{align}
In terms of the electronic eigenstates, $\ket*{\Psi_\Gamma^\mu} = \sum_\lambda (U_\Gamma)_{\lambda\mu} \ket*{\phi_\Gamma^\lambda}$,
$\hat{H}_\mathrm{el-ph}$ is written as
\begin{align}
   & \hat{H}_\mathrm{el-ph}
  = \sum_\Gamma \hat{u}_{-\Gamma} \hat{W}_\Gamma,
  \label{eq:Hel_append1}
  \\
   & \hat{W}_\Gamma
  = \sum_{\gamma,\mu\mu'} V_{\gamma\mu,\gamma+\Gamma\mu'}
  \ketbra*{\Psi_\gamma^\mu}{\Psi_{\gamma+\Gamma}^{\mu'}},
  \label{eq:Hel_append2}
\end{align}
where the coupling constant is defined by
\begin{align}
  V_{\gamma\mu,\gamma+\Gamma\mu'}
   & = \sum_{\ell\lambda\nu s} \frac{\zeta}{\sqrt{3N}} t_{1+\ell,1}^{\lambda\nu}
  e^{i\ell\Phi_\gamma^\lambda} e^{i\ell m_s\alpha/2}( 1 - e^{i\ell\Omega_\Gamma^s} )
  \notag                                                                         \\
   & \quad \times
  q_\ell^{(s)} (U_\gamma)_{\lambda\mu}^\ast (U_{\gamma+\Gamma})_{\nu\mu'}
  (v_\Gamma^{(s)})^\ast.
  \label{eq:Hel_append3}
\end{align}
The coupling constant can be expressed as the inner product between the phonon polarization vector and the coupling vector, $\bm{v}_\Gamma \cdot \bm{V}_{\gamma\mu,\gamma+\Gamma\mu'}$,
where the latter is given by
\begin{align}
  \bm{V}_{\gamma\mu,\gamma+\Gamma\mu'}
   & = \sum_{\ell\lambda\nu s} \frac{\zeta}{\sqrt{3N}} t_{1+\ell,1}^{\lambda\nu}
  e^{i\ell\Phi_\gamma^\lambda} e^{i\ell m_s\alpha/2}( 1 - e^{i\ell\Omega_\Gamma^s} )
  \notag                                                                         \\
   & \quad \times
  q_\ell^{(s)} (U_\gamma)_{\lambda\mu}^\ast (U_{\gamma+\Gamma})_{\nu\mu'}
  \bm{e}^{(s)}.
\end{align}

\subsection{Mechanical angular momentum of phonons}~\label{subsec:Lph}

The phonon mechanical angular momentum (MAM) operator along the screw axis (the $z$ axis) is defined as
\begin{equation}
  \hat{L}_z
  = \sum_{\ell=1}^{3N} (\hat{\bm{u}}_\ell \times \hat{\bm{p}}_\ell )_z,
\end{equation}
where $\hat{\bm{p}}_\ell$ is the canonical momentum operator at the $\ell$th atomic site.
In the circular basis, this is rewritten as
\begin{equation}
  \hat{L}_z
  = i \sum_{\ell=1}^{3N}
  \ab( \hat{u}_\ell^{(+)} \hat{p}_\ell^{(-)} - \hat{u}_\ell^{(-)} \hat{p}_\ell^{(+)} ).
  \label{eq:Lz_chiral_basis}
\end{equation}
To evaluate the MAM in the thermal equilibrium state,
$\hat{u}_\ell^{(s)}$ and $\hat{p}_\ell^{(s)}$ are decomposed into the symmetry-adapted basis,
the former of which is already given by
\begin{equation}
  \hat{u}_\ell^{(s)}
  = \frac{1}{\sqrt{3N}} \sum_{\Gamma} e^{-i(\ell-1)\Omega_{\Gamma}^{(s)}}
  \ab( \frac{\hbar}{2M\omega_\Gamma} )^{1/2}
  ( \hat{a}_\Gamma + \hat{a}_{-\Gamma}^\dagger ) v_\Gamma^{(s)},
\end{equation}
and the latter is,
by making use of $\hat{p}_\ell^{(s)} = M \dot{\hat{u}}_\ell^{(s)}$ and $M\dot{\hat{u}}_\Gamma=\hat{p}_{-\Gamma}$, obtained as
\begin{equation}
  \hat{p}_\ell^{(s)}
  = \frac{1}{\sqrt{3N}} \sum_{\Gamma} e^{-i(\ell-1)\Omega_{-\Gamma}^{(s)}}
  i \ab( \frac{\hbar M\omega_\Gamma}{2} )^{1/2}
  ( \hat{a}_\Gamma^\dagger - \hat{a}_{-\Gamma} ) v_{-\Gamma}^{(s)}.
\end{equation}
Substituting these expressions yields
\begin{equation}
  \aab*{\hat{L}_z}
  = \sum_\Gamma \hbar \ab( \vab*{v_\Gamma^{(-)}}^2 - \vab*{v_\Gamma^{(+)}}^2 )
  \ab( f(\omega_\Gamma) + \frac{1}{2} ).
  \label{eq:phonon_mam}
\end{equation}
where the time-reversal symmetry relation $v_{-\Gamma}^{(-s)} = (v_\Gamma^{(s)})^\ast$ is used.
This is the thermal average of the phonon spin angular momentum of each irrep.~$\Gamma$,
because the quantity $\vab*{v_\Gamma^{(-)}}^2 - \vab*{v_\Gamma^{(+)}}^2 = -\sum_s m_s \vab*{v_\Gamma^{(s)}}^2$ is the mean value of the spin angular momentum of each eigenvector.

This procedure cannot be generalized to $x$ and $y$ components,
because the above derivation explicitly relies on the circular basis along $z$ axis.
Nevertheless, it is always possible to express the mean spin angular momentum along the $a$ axis ($a=x,y,z$) as $-i\sum_{b,c=x,y,z} \epsilon_{abc} (v_\Gamma^b)^\ast v_\Gamma^c$.
Therefore, we can extend the definition of the phonon MAM to
\begin{align}
  \aab*{\hat{\bm{L}}}
   & = \sum_\Gamma L_{\Gamma} \ab( f(\omega_\Gamma) + \frac{1}{2} ),
  \\
  \bm{L}_{\Gamma}
   & = -i\hbar \bm{v}_\Gamma^\ast \times \bm{v}_\Gamma
  = -i\hbar \bm{v}_{-\Gamma} \times \bm{v}_\Gamma.
\end{align}

\subsection{Orbital angular momentum of Bloch electrons}~\label{subsec:Lel}

Orbital angular momentum (OAM) operator of electrons is defined by
\begin{equation}
  \hat{\bm{L}}_\mathrm{el}
  = \frac{m_\mathrm{e}}{2}
  \ab( \hat{\bm{r}} \times \hat{\bm{v}} - \hat{\bm{v}} \times \hat{\bm{r}} ),
\end{equation}
where $m_\mathrm{e}$ is the electronic mass,
$\hat{\bm{r}}$ is the position operator, and
$\hat{\bm{v}} = (i/\hbar)[\hat{H}_\mathrm{el},\hat{\bm{r}}]$ is the velocity operator.
The matrix element of the $z$-component of OAM becomes
\begin{align}
  \braket*[3]{\Psi_\Gamma^\mu}{\hat{L}_{\mathrm{el},z}}{\Psi_\Gamma^\mu}
   & = -\frac{im_\mathrm{e}}{\hbar} \sum_{b,c} \epsilon_{zbc}
  \left[ E_{\Gamma\mu} \braket*[3]{\Psi_\Gamma^\mu}{\hat{r}_b\hat{r}_c}{\Psi_\Gamma^\mu}
    \right.
  \notag                                                      \\
   & \quad \left.
    - \braket*[3]{\Psi_\Gamma^\mu}{\hat{r}_b\hat{H}_\mathrm{el}\hat{r}_c}{\Psi_\Gamma^\mu} \right].
  \label{eq:OAM_z_Bloch}
\end{align}
Rewriting in terms of $\ket*{\phi_\ell^\lambda}$
and assuming that the atomic orbital state $\ket*{\phi_\ell^\lambda}$ is spatially localized, and hence, is the eigenstate of the position operator as $\hat{\bm{r}} \ket*{\phi_\ell^\lambda} = \bm{R}_\ell \ket*{\phi_\ell^\lambda}$, the first term of the right-hand-side of above equation yields
\begin{equation}
  \braket*[3]{\Psi_\Gamma^\mu}{\hat{r}_b\hat{r}_c}{\Psi_\Gamma^\mu}
  = \frac{1}{3N} \sum_\ell R_{\ell,b} R_{\ell,c},
\end{equation}
where the unitary condition of $U_\Gamma$ is used.
This term does not contribute to OAM, because the equality $\sum_{b,c}\epsilon_{abc} R_{\ell,b} R_{\ell,c} = 0$ is satisfied.
In the similar manner, the second term of Eq.~\eqref{eq:OAM_z_Bloch} is rewritten as
\begin{align}
  \braket*[3]{\Psi_\Gamma^\mu}{\hat{r}_b\hat{H}_\mathrm{el}\hat{r}_c}{\Psi_\Gamma^\mu}
   & = \sum_\ell \sum_{\lambda,\lambda'}
  (U_\Gamma)_{\lambda'\mu}^\ast (U_\Gamma)_{\lambda\mu}
  e^{i\ell\Phi_\Gamma^{\lambda'}}
  \notag                                 \\
   & \quad \times
  w_{bc}^\ell
  \braket*[3]{\phi_{1+\ell}^{\lambda'}}{\hat{H}_\mathrm{el}}{\phi_1^\lambda},
\end{align}
with $w_{bc}^\ell = \sum_{\ell'} R_{\ell+\ell',b} R_{\ell',c} /3N$.
Using $\sum_{bc}\epsilon_{zbc} w_{bc}^\ell = -\rho^2 \sin(\ell\alpha)$,
the OAM is finally expressed as
\begin{align}
  \braket*[3]{\Psi_\Gamma^\mu}{\hat{L}_{\mathrm{el},z}}{\Psi_\Gamma^\mu}
   & = -\frac{im_\mathrm{e}\rho^2}{\hbar} \sum_\ell \sum_{\lambda,\lambda'}
  (U_\Gamma)_{\lambda'\mu}^\ast (U_\Gamma)_{\lambda\mu}
  e^{i\ell\Phi_\Gamma^{\lambda'}}
  \notag                                                                    \\
   & \quad \times \sin(\ell\alpha)
  \braket*[3]{\phi_{1+\ell}^{\lambda'}}{\hat{H}_\mathrm{el}}{\phi_1^\lambda}.
  \label{eq:Lz_general}
\end{align}
This is rewritten as
\begin{equation}
  \braket*[3]{\Psi_\Gamma^\mu}{\hat{L}_{\mathrm{el},z}}{\Psi_\Gamma^\mu}
  = -m_\mathrm{e} \sum_{\ell,\lambda,\lambda'}
  (\bm{R}_{1+\ell} \times \bm{R}_1 )_z J_{\ell;\Gamma\mu}^{\lambda'\lambda},
\end{equation}
where $(\bm{R}_{1+\ell} \times \bm{R}_1 )_z = -\rho^2\sin(\ell\alpha)$ is the area spanned by the vectors associated with the hopping from site $1$ to $1+\ell$, and
\begin{equation}
  J_{\ell;\Gamma\mu}^{\lambda'\lambda}
  = -\frac{i}{\hbar} (U_\Gamma)_{\lambda'\mu}^\ast (U_\Gamma)_{\lambda\mu}
  e^{i\ell\Phi_\Gamma^{\lambda'}} \braket*[3]{\phi_{1+\ell}^{\lambda'}}{\hat{H}_\mathrm{el}}{\phi_1^\lambda},
\end{equation}
is the corresponding hopping rate of $\ket*{\Psi_\Gamma^\mu}$ from orbital $\lambda$ to $\lambda'$.
Therefore, OAM can be interpreted as the sum of the hopping rate weighed by the hopping area.

When we include the contribution from electron--phonon interaction, $\hat{H}_\mathrm{el}$ is replaced by $\hat{H}_\mathrm{el}'=\hat{H}_\mathrm{el}+\hat{H}_2$ in Eq.~\eqref{eq:Lz_general}, which indicates that the electron-phonon interaction directly affect the OAM.

\bibliography{ref}

\begin{thebibliography}{42}%
\makeatletter
\providecommand \@ifxundefined [1]{%
 \@ifx{#1\undefined}
}%
\providecommand \@ifnum [1]{%
 \ifnum #1\expandafter \@firstoftwo
 \else \expandafter \@secondoftwo
 \fi
}%
\providecommand \@ifx [1]{%
 \ifx #1\expandafter \@firstoftwo
 \else \expandafter \@secondoftwo
 \fi
}%
\providecommand \natexlab [1]{#1}%
\providecommand \enquote  [1]{``#1''}%
\providecommand \bibnamefont  [1]{#1}%
\providecommand \bibfnamefont [1]{#1}%
\providecommand \citenamefont [1]{#1}%
\providecommand \href@noop [0]{\@secondoftwo}%
\providecommand \href [0]{\begingroup \@sanitize@url \@href}%
\providecommand \@href[1]{\@@startlink{#1}\@@href}%
\providecommand \@@href[1]{\endgroup#1\@@endlink}%
\providecommand \@sanitize@url [0]{\catcode `\\12\catcode `\$12\catcode `\&12\catcode `\#12\catcode `\^12\catcode `\_12\catcode `\%12\relax}%
\providecommand \@@startlink[1]{}%
\providecommand \@@endlink[0]{}%
\providecommand \url  [0]{\begingroup\@sanitize@url \@url }%
\providecommand \@url [1]{\endgroup\@href {#1}{\urlprefix }}%
\providecommand \urlprefix  [0]{URL }%
\providecommand \Eprint [0]{\href }%
\providecommand \doibase [0]{https://doi.org/}%
\providecommand \selectlanguage [0]{\@gobble}%
\providecommand \bibinfo  [0]{\@secondoftwo}%
\providecommand \bibfield  [0]{\@secondoftwo}%
\providecommand \translation [1]{[#1]}%
\providecommand \BibitemOpen [0]{}%
\providecommand \bibitemStop [0]{}%
\providecommand \bibitemNoStop [0]{.\EOS\space}%
\providecommand \EOS [0]{\spacefactor3000\relax}%
\providecommand \BibitemShut  [1]{\csname bibitem#1\endcsname}%
\let\auto@bib@innerbib\@empty
\bibitem [{\citenamefont {Zhang}\ and\ \citenamefont {Niu}(2015)}]{zhang2015Chiral}%
  \BibitemOpen
  \bibfield  {author} {\bibinfo {author} {\bibfnamefont {L.}~\bibnamefont {Zhang}}\ and\ \bibinfo {author} {\bibfnamefont {Q.}~\bibnamefont {Niu}},\ }\bibfield  {title} {\bibinfo {title} {Chiral {{Phonons}} at {{High-Symmetry Points}} in {{Monolayer Hexagonal Lattices}}},\ }\href {https://doi.org/10.1103/PhysRevLett.115.115502} {\bibfield  {journal} {\bibinfo  {journal} {Phys. Rev. Lett.}\ }\textbf {\bibinfo {volume} {115}},\ \bibinfo {pages} {115502} (\bibinfo {year} {2015})}\BibitemShut {NoStop}%
\bibitem [{\citenamefont {Zhang}\ and\ \citenamefont {Murakami}(2022)}]{zhang2022Chiral}%
  \BibitemOpen
  \bibfield  {author} {\bibinfo {author} {\bibfnamefont {T.}~\bibnamefont {Zhang}}\ and\ \bibinfo {author} {\bibfnamefont {S.}~\bibnamefont {Murakami}},\ }\bibfield  {title} {\bibinfo {title} {Chiral phonons and pseudoangular momentum in nonsymmorphic systems},\ }\href {https://doi.org/10.1103/PhysRevResearch.4.L012024} {\bibfield  {journal} {\bibinfo  {journal} {Phys. Rev. Res.}\ }\textbf {\bibinfo {volume} {4}},\ \bibinfo {pages} {L012024} (\bibinfo {year} {2022})}\BibitemShut {NoStop}%
\bibitem [{\citenamefont {Tsunetsugu}\ and\ \citenamefont {Kusunose}(2023)}]{tsunetsugu2023Theory}%
  \BibitemOpen
  \bibfield  {author} {\bibinfo {author} {\bibfnamefont {H.}~\bibnamefont {Tsunetsugu}}\ and\ \bibinfo {author} {\bibfnamefont {H.}~\bibnamefont {Kusunose}},\ }\bibfield  {title} {\bibinfo {title} {Theory of {{Energy Dispersion}} of {{Chiral Phonons}}},\ }\href {https://doi.org/10.7566/JPSJ.92.023601} {\bibfield  {journal} {\bibinfo  {journal} {J. Phys. Soc. Jpn.}\ }\textbf {\bibinfo {volume} {92}},\ \bibinfo {pages} {023601} (\bibinfo {year} {2023})}\BibitemShut {NoStop}%
\bibitem [{\citenamefont {Kato}\ and\ \citenamefont {Kishine}(2023)}]{kato2023Note}%
  \BibitemOpen
  \bibfield  {author} {\bibinfo {author} {\bibfnamefont {A.}~\bibnamefont {Kato}}\ and\ \bibinfo {author} {\bibfnamefont {J.}~\bibnamefont {Kishine}},\ }\bibfield  {title} {\bibinfo {title} {Note on {{Angular Momentum}} of {{Phonons}} in {{Chiral Crystals}}},\ }\href {https://doi.org/10.7566/JPSJ.92.075002} {\bibfield  {journal} {\bibinfo  {journal} {J. Phys. Soc. Jpn.}\ }\textbf {\bibinfo {volume} {92}},\ \bibinfo {pages} {075002} (\bibinfo {year} {2023})}\BibitemShut {NoStop}%
\bibitem [{\citenamefont {Ishito}\ \emph {et~al.}(2022)\citenamefont {Ishito}, \citenamefont {Mao}, \citenamefont {Kousaka}, \citenamefont {Togawa}, \citenamefont {Iwasaki}, \citenamefont {Zhang}, \citenamefont {Murakami}, \citenamefont {Kishine},\ and\ \citenamefont {Satoh}}]{ishito2022Truly}%
  \BibitemOpen
  \bibfield  {author} {\bibinfo {author} {\bibfnamefont {K.}~\bibnamefont {Ishito}}, \bibinfo {author} {\bibfnamefont {H.}~\bibnamefont {Mao}}, \bibinfo {author} {\bibfnamefont {Y.}~\bibnamefont {Kousaka}}, \bibinfo {author} {\bibfnamefont {Y.}~\bibnamefont {Togawa}}, \bibinfo {author} {\bibfnamefont {S.}~\bibnamefont {Iwasaki}}, \bibinfo {author} {\bibfnamefont {T.}~\bibnamefont {Zhang}}, \bibinfo {author} {\bibfnamefont {S.}~\bibnamefont {Murakami}}, \bibinfo {author} {\bibfnamefont {J.}~\bibnamefont {Kishine}},\ and\ \bibinfo {author} {\bibfnamefont {T.}~\bibnamefont {Satoh}},\ }\bibfield  {title} {\bibinfo {title} {Truly chiral phonons in {$\alpha$}-{{HgS}}},\ }\href {https://doi.org/10.1038/s41567-022-01790-x} {\bibfield  {journal} {\bibinfo  {journal} {Nat. Phys.}\ }\textbf {\bibinfo {volume} {19}},\ \bibinfo {pages} {35} (\bibinfo {year} {2022})}\BibitemShut {NoStop}%
\bibitem [{\citenamefont {Ishito}\ \emph {et~al.}(2023)\citenamefont {Ishito}, \citenamefont {Mao}, \citenamefont {Kobayashi}, \citenamefont {Kousaka}, \citenamefont {Togawa}, \citenamefont {Kusunose}, \citenamefont {Kishine},\ and\ \citenamefont {Satoh}}]{ishito2023Chiral}%
  \BibitemOpen
  \bibfield  {author} {\bibinfo {author} {\bibfnamefont {K.}~\bibnamefont {Ishito}}, \bibinfo {author} {\bibfnamefont {H.}~\bibnamefont {Mao}}, \bibinfo {author} {\bibfnamefont {K.}~\bibnamefont {Kobayashi}}, \bibinfo {author} {\bibfnamefont {Y.}~\bibnamefont {Kousaka}}, \bibinfo {author} {\bibfnamefont {Y.}~\bibnamefont {Togawa}}, \bibinfo {author} {\bibfnamefont {H.}~\bibnamefont {Kusunose}}, \bibinfo {author} {\bibfnamefont {J.}~\bibnamefont {Kishine}},\ and\ \bibinfo {author} {\bibfnamefont {T.}~\bibnamefont {Satoh}},\ }\bibfield  {title} {\bibinfo {title} {Chiral phonons: Circularly polarized {{Raman}} spectroscopy and ab initio calculations in a chiral crystal tellurium},\ }\href {https://doi.org/10.1002/chir.23544} {\bibfield  {journal} {\bibinfo  {journal} {Chirality}\ }\textbf {\bibinfo {volume} {35}},\ \bibinfo {pages} {338} (\bibinfo {year} {2023})}\BibitemShut {NoStop}%
\bibitem [{\citenamefont {Oishi}\ \emph {et~al.}(2024)\citenamefont {Oishi}, \citenamefont {Fujii},\ and\ \citenamefont {Koreeda}}]{oishi2024Selective}%
  \BibitemOpen
  \bibfield  {author} {\bibinfo {author} {\bibfnamefont {E.}~\bibnamefont {Oishi}}, \bibinfo {author} {\bibfnamefont {Y.}~\bibnamefont {Fujii}},\ and\ \bibinfo {author} {\bibfnamefont {A.}~\bibnamefont {Koreeda}},\ }\bibfield  {title} {\bibinfo {title} {Selective observation of enantiomeric chiral phonons in $\alpha$-quartz},\ }\href {https://doi.org/10.1103/PhysRevB.109.104306} {\bibfield  {journal} {\bibinfo  {journal} {Phys. Rev. B}\ }\textbf {\bibinfo {volume} {109}},\ \bibinfo {pages} {104306} (\bibinfo {year} {2024})}\BibitemShut {NoStop}%
\bibitem [{\citenamefont {Ueda}\ \emph {et~al.}(2023)\citenamefont {Ueda}, \citenamefont {{Garc{\'i}a-Fern{\'a}ndez}}, \citenamefont {Agrestini}, \citenamefont {Romao}, \citenamefont {{van den Brink}}, \citenamefont {Spaldin}, \citenamefont {Zhou},\ and\ \citenamefont {Staub}}]{ueda2023Chiral}%
  \BibitemOpen
  \bibfield  {author} {\bibinfo {author} {\bibfnamefont {H.}~\bibnamefont {Ueda}}, \bibinfo {author} {\bibfnamefont {M.}~\bibnamefont {{Garc{\'i}a-Fern{\'a}ndez}}}, \bibinfo {author} {\bibfnamefont {S.}~\bibnamefont {Agrestini}}, \bibinfo {author} {\bibfnamefont {C.~P.}\ \bibnamefont {Romao}}, \bibinfo {author} {\bibfnamefont {J.}~\bibnamefont {{van den Brink}}}, \bibinfo {author} {\bibfnamefont {N.~A.}\ \bibnamefont {Spaldin}}, \bibinfo {author} {\bibfnamefont {K.-J.}\ \bibnamefont {Zhou}},\ and\ \bibinfo {author} {\bibfnamefont {U.}~\bibnamefont {Staub}},\ }\bibfield  {title} {\bibinfo {title} {Chiral phonons in quartz probed by {{X-rays}}},\ }\href {https://doi.org/10.1038/s41586-023-06016-5} {\bibfield  {journal} {\bibinfo  {journal} {Nature}\ }\textbf {\bibinfo {volume} {618}},\ \bibinfo {pages} {946} (\bibinfo {year} {2023})}\BibitemShut {NoStop}%
\bibitem [{\citenamefont {Bo{\v z}ovic}(1984)}]{bozovic1984Possible}%
  \BibitemOpen
  \bibfield  {author} {\bibinfo {author} {\bibfnamefont {I.}~\bibnamefont {Bo{\v z}ovic}},\ }\bibfield  {title} {\bibinfo {title} {Possible band-structure shapes of quasi-one-dimensional solids},\ }\href {https://doi.org/10.1103/PhysRevB.29.6586} {\bibfield  {journal} {\bibinfo  {journal} {Phys. Rev. B}\ }\textbf {\bibinfo {volume} {29}},\ \bibinfo {pages} {6586} (\bibinfo {year} {1984})}\BibitemShut {NoStop}%
\bibitem [{\citenamefont {Zhang}\ and\ \citenamefont {Niu}(2014)}]{zhang2014Angular}%
  \BibitemOpen
  \bibfield  {author} {\bibinfo {author} {\bibfnamefont {L.}~\bibnamefont {Zhang}}\ and\ \bibinfo {author} {\bibfnamefont {Q.}~\bibnamefont {Niu}},\ }\bibfield  {title} {\bibinfo {title} {Angular {{Momentum}} of {{Phonons}} and the {{Einstein--de Haas Effect}}},\ }\href {https://doi.org/10.1103/PhysRevLett.112.085503} {\bibfield  {journal} {\bibinfo  {journal} {Phys. Rev. Lett.}\ }\textbf {\bibinfo {volume} {112}},\ \bibinfo {pages} {085503} (\bibinfo {year} {2014})}\BibitemShut {NoStop}%
\bibitem [{\citenamefont {Vonsovskii}\ and\ \citenamefont {Svirskii}(1962)}]{vonsovskii1962Phonon}%
  \BibitemOpen
  \bibfield  {author} {\bibinfo {author} {\bibfnamefont {S.~V.}\ \bibnamefont {Vonsovskii}}\ and\ \bibinfo {author} {\bibfnamefont {M.~S.}\ \bibnamefont {Svirskii}},\ }\bibfield  {title} {\bibinfo {title} {Phonon {{Spin}}},\ }\href@noop {} {\bibfield  {journal} {\bibinfo  {journal} {Sov. Phys. Solid State}\ }\textbf {\bibinfo {volume} {3}},\ \bibinfo {pages} {1568} (\bibinfo {year} {1962})}\BibitemShut {NoStop}%
\bibitem [{\citenamefont {Bloom}\ \emph {et~al.}(2024)\citenamefont {Bloom}, \citenamefont {Paltiel}, \citenamefont {Naaman},\ and\ \citenamefont {Waldeck}}]{bloom2024Chiral}%
  \BibitemOpen
  \bibfield  {author} {\bibinfo {author} {\bibfnamefont {B.~P.}\ \bibnamefont {Bloom}}, \bibinfo {author} {\bibfnamefont {Y.}~\bibnamefont {Paltiel}}, \bibinfo {author} {\bibfnamefont {R.}~\bibnamefont {Naaman}},\ and\ \bibinfo {author} {\bibfnamefont {D.~H.}\ \bibnamefont {Waldeck}},\ }\bibfield  {title} {\bibinfo {title} {Chiral {{Induced Spin Selectivity}}},\ }\href {https://doi.org/10.1021/acs.chemrev.3c00661} {\bibfield  {journal} {\bibinfo  {journal} {Chem. Rev.}\ }\textbf {\bibinfo {volume} {124}},\ \bibinfo {pages} {1950} (\bibinfo {year} {2024})}\BibitemShut {NoStop}%
\bibitem [{\citenamefont {Fransson}(2020)}]{fransson2020Vibrational}%
  \BibitemOpen
  \bibfield  {author} {\bibinfo {author} {\bibfnamefont {J.}~\bibnamefont {Fransson}},\ }\bibfield  {title} {\bibinfo {title} {Vibrational origin of exchange splitting and ''chiral-induced spin selectivity},\ }\href {https://doi.org/10.1103/PhysRevB.102.235416} {\bibfield  {journal} {\bibinfo  {journal} {Phys. Rev. B}\ }\textbf {\bibinfo {volume} {102}},\ \bibinfo {pages} {235416} (\bibinfo {year} {2020})}\BibitemShut {NoStop}%
\bibitem [{\citenamefont {Du}\ \emph {et~al.}(2020)\citenamefont {Du}, \citenamefont {Fu},\ and\ \citenamefont {Wu}}]{du2020Vibrationenhanced}%
  \BibitemOpen
  \bibfield  {author} {\bibinfo {author} {\bibfnamefont {G.-F.}\ \bibnamefont {Du}}, \bibinfo {author} {\bibfnamefont {H.-H.}\ \bibnamefont {Fu}},\ and\ \bibinfo {author} {\bibfnamefont {R.}~\bibnamefont {Wu}},\ }\bibfield  {title} {\bibinfo {title} {Vibration-enhanced spin-selective transport of electrons in the {{DNA}} double helix},\ }\href {https://doi.org/10.1103/PhysRevB.102.035431} {\bibfield  {journal} {\bibinfo  {journal} {Phys. Rev. B}\ }\textbf {\bibinfo {volume} {102}},\ \bibinfo {pages} {035431} (\bibinfo {year} {2020})}\BibitemShut {NoStop}%
\bibitem [{\citenamefont {Fransson}(2023)}]{fransson2023Chiral}%
  \BibitemOpen
  \bibfield  {author} {\bibinfo {author} {\bibfnamefont {J.}~\bibnamefont {Fransson}},\ }\bibfield  {title} {\bibinfo {title} {Chiral phonon induced spin polarization},\ }\href {https://doi.org/10.1103/PhysRevResearch.5.L022039} {\bibfield  {journal} {\bibinfo  {journal} {Phys. Rev. Res.}\ }\textbf {\bibinfo {volume} {5}},\ \bibinfo {pages} {L022039} (\bibinfo {year} {2023})}\BibitemShut {NoStop}%
\bibitem [{\citenamefont {Funato}\ \emph {et~al.}(2024)\citenamefont {Funato}, \citenamefont {Matsuo},\ and\ \citenamefont {Kato}}]{funato2024ChiralityInduced}%
  \BibitemOpen
  \bibfield  {author} {\bibinfo {author} {\bibfnamefont {T.}~\bibnamefont {Funato}}, \bibinfo {author} {\bibfnamefont {M.}~\bibnamefont {Matsuo}},\ and\ \bibinfo {author} {\bibfnamefont {T.}~\bibnamefont {Kato}},\ }\bibfield  {title} {\bibinfo {title} {Chirality-{{Induced Phonon-Spin Conversion}} at an {{Interface}}},\ }\href {https://doi.org/10.1103/PhysRevLett.132.236201} {\bibfield  {journal} {\bibinfo  {journal} {Phys. Rev. Lett.}\ }\textbf {\bibinfo {volume} {132}},\ \bibinfo {pages} {236201} (\bibinfo {year} {2024})}\BibitemShut {NoStop}%
\bibitem [{\citenamefont {Inui}\ \emph {et~al.}(2020)\citenamefont {Inui}, \citenamefont {Aoki}, \citenamefont {Nishiue}, \citenamefont {Shiota}, \citenamefont {Kousaka}, \citenamefont {Shishido}, \citenamefont {Hirobe}, \citenamefont {Suda}, \citenamefont {Ohe}, \citenamefont {Kishine}, \citenamefont {Yamamoto},\ and\ \citenamefont {Togawa}}]{inui2020ChiralityInduced}%
  \BibitemOpen
  \bibfield  {author} {\bibinfo {author} {\bibfnamefont {A.}~\bibnamefont {Inui}}, \bibinfo {author} {\bibfnamefont {R.}~\bibnamefont {Aoki}}, \bibinfo {author} {\bibfnamefont {Y.}~\bibnamefont {Nishiue}}, \bibinfo {author} {\bibfnamefont {K.}~\bibnamefont {Shiota}}, \bibinfo {author} {\bibfnamefont {Y.}~\bibnamefont {Kousaka}}, \bibinfo {author} {\bibfnamefont {H.}~\bibnamefont {Shishido}}, \bibinfo {author} {\bibfnamefont {D.}~\bibnamefont {Hirobe}}, \bibinfo {author} {\bibfnamefont {M.}~\bibnamefont {Suda}}, \bibinfo {author} {\bibfnamefont {J.}~\bibnamefont {Ohe}}, \bibinfo {author} {\bibfnamefont {J.}~\bibnamefont {Kishine}}, \bibinfo {author} {\bibfnamefont {H.~M.}\ \bibnamefont {Yamamoto}},\ and\ \bibinfo {author} {\bibfnamefont {Y.}~\bibnamefont {Togawa}},\ }\bibfield  {title} {\bibinfo {title} {Chirality-{{Induced Spin-Polarized State}} of a {{Chiral Crystal}} $\mathrm{CrNb_3S_6}$},\ }\href {https://doi.org/10.1103/PhysRevLett.124.166602} {\bibfield  {journal} {\bibinfo  {journal} {Phys. Rev. Lett.}\ }\textbf {\bibinfo {volume} {124}},\ \bibinfo {pages} {166602} (\bibinfo {year} {2020})}\BibitemShut {NoStop}%
\bibitem [{\citenamefont {Rikken}\ \emph {et~al.}(2002)\citenamefont {Rikken}, \citenamefont {Strohm},\ and\ \citenamefont {Wyder}}]{rikken2002Observation}%
  \BibitemOpen
  \bibfield  {author} {\bibinfo {author} {\bibfnamefont {G.~L. J.~A.}\ \bibnamefont {Rikken}}, \bibinfo {author} {\bibfnamefont {C.}~\bibnamefont {Strohm}},\ and\ \bibinfo {author} {\bibfnamefont {P.}~\bibnamefont {Wyder}},\ }\bibfield  {title} {\bibinfo {title} {Observation of {{Magnetoelectric Directional Anisotropy}}},\ }\href {https://doi.org/10.1103/PhysRevLett.89.133005} {\bibfield  {journal} {\bibinfo  {journal} {Phys. Rev. Lett.}\ }\textbf {\bibinfo {volume} {89}},\ \bibinfo {pages} {133005} (\bibinfo {year} {2002})}\BibitemShut {NoStop}%
\bibitem [{\citenamefont {Furukawa}\ \emph {et~al.}(2017)\citenamefont {Furukawa}, \citenamefont {Shimokawa}, \citenamefont {Kobayashi},\ and\ \citenamefont {Itou}}]{furukawa2017Observation}%
  \BibitemOpen
  \bibfield  {author} {\bibinfo {author} {\bibfnamefont {T.}~\bibnamefont {Furukawa}}, \bibinfo {author} {\bibfnamefont {Y.}~\bibnamefont {Shimokawa}}, \bibinfo {author} {\bibfnamefont {K.}~\bibnamefont {Kobayashi}},\ and\ \bibinfo {author} {\bibfnamefont {T.}~\bibnamefont {Itou}},\ }\bibfield  {title} {\bibinfo {title} {Observation of current-induced bulk magnetization in elemental tellurium},\ }\href {https://doi.org/10.1038/s41467-017-01093-3} {\bibfield  {journal} {\bibinfo  {journal} {Nat. Commun.}\ }\textbf {\bibinfo {volume} {8}},\ \bibinfo {pages} {954} (\bibinfo {year} {2017})}\BibitemShut {NoStop}%
\bibitem [{\citenamefont {Ohe}\ \emph {et~al.}(2024)\citenamefont {Ohe}, \citenamefont {Shishido}, \citenamefont {Kato}, \citenamefont {Utsumi}, \citenamefont {Matsuura},\ and\ \citenamefont {Togawa}}]{ohe2024ChiralityInduced}%
  \BibitemOpen
  \bibfield  {author} {\bibinfo {author} {\bibfnamefont {K.}~\bibnamefont {Ohe}}, \bibinfo {author} {\bibfnamefont {H.}~\bibnamefont {Shishido}}, \bibinfo {author} {\bibfnamefont {M.}~\bibnamefont {Kato}}, \bibinfo {author} {\bibfnamefont {S.}~\bibnamefont {Utsumi}}, \bibinfo {author} {\bibfnamefont {H.}~\bibnamefont {Matsuura}},\ and\ \bibinfo {author} {\bibfnamefont {Y.}~\bibnamefont {Togawa}},\ }\bibfield  {title} {\bibinfo {title} {Chirality-{{Induced Selectivity}} of {{Phonon Angular Momenta}} in {{Chiral Quartz Crystals}}},\ }\href {https://doi.org/10.1103/PhysRevLett.132.056302} {\bibfield  {journal} {\bibinfo  {journal} {Phys. Rev. Lett.}\ }\textbf {\bibinfo {volume} {132}},\ \bibinfo {pages} {056302} (\bibinfo {year} {2024})}\BibitemShut {NoStop}%
\bibitem [{\citenamefont {Bousquet}\ \emph {et~al.}(2025)\citenamefont {Bousquet}, \citenamefont {Fava}, \citenamefont {Romestan}, \citenamefont {{G{\'o}mez-Ortiz}}, \citenamefont {McCabe},\ and\ \citenamefont {Romero}}]{bousquet2025Structural}%
  \BibitemOpen
  \bibfield  {author} {\bibinfo {author} {\bibfnamefont {E.}~\bibnamefont {Bousquet}}, \bibinfo {author} {\bibfnamefont {M.}~\bibnamefont {Fava}}, \bibinfo {author} {\bibfnamefont {Z.}~\bibnamefont {Romestan}}, \bibinfo {author} {\bibfnamefont {F.}~\bibnamefont {{G{\'o}mez-Ortiz}}}, \bibinfo {author} {\bibfnamefont {E.~E.}\ \bibnamefont {McCabe}},\ and\ \bibinfo {author} {\bibfnamefont {A.~H.}\ \bibnamefont {Romero}},\ }\bibfield  {title} {\bibinfo {title} {Structural chirality and related properties in periodic inorganic solids: Review and perspectives},\ }\href {https://doi.org/10.1088/1361-648X/adb674} {\bibfield  {journal} {\bibinfo  {journal} {J. Phys.: Condens. Matter}\ }\textbf {\bibinfo {volume} {37}},\ \bibinfo {pages} {163004} (\bibinfo {year} {2025})}\BibitemShut {NoStop}%
\bibitem [{\citenamefont {Tateishi}\ \emph {et~al.}(2025)\citenamefont {Tateishi}, \citenamefont {Kato},\ and\ \citenamefont {Kishine}}]{tateishi2025Electron}%
  \BibitemOpen
  \bibfield  {author} {\bibinfo {author} {\bibfnamefont {T.}~\bibnamefont {Tateishi}}, \bibinfo {author} {\bibfnamefont {A.}~\bibnamefont {Kato}},\ and\ \bibinfo {author} {\bibfnamefont {J.}~\bibnamefont {Kishine}},\ }\bibfield  {title} {\bibinfo {title} {Electron--{{Chiral Phonon Coupling}}, {{Crystal Angular Momentum}}, and {{Phonon Chirality}}},\ }\href {https://doi.org/10.7566/JPSJ.94.053601} {\bibfield  {journal} {\bibinfo  {journal} {J. Phys. Soc. Jpn.}\ }\textbf {\bibinfo {volume} {94}},\ \bibinfo {pages} {053601} (\bibinfo {year} {2025})}\BibitemShut {NoStop}%
\bibitem [{\citenamefont {Tateishi}\ \emph {et~al.}(2026)\citenamefont {Tateishi}, \citenamefont {Kato}, \citenamefont {Ovchinnikov},\ and\ \citenamefont {Kishine}}]{tateishi2026Microscopic}%
  \BibitemOpen
  \bibfield  {author} {\bibinfo {author} {\bibfnamefont {T.}~\bibnamefont {Tateishi}}, \bibinfo {author} {\bibfnamefont {A.}~\bibnamefont {Kato}}, \bibinfo {author} {\bibfnamefont {A.~S.}\ \bibnamefont {Ovchinnikov}},\ and\ \bibinfo {author} {\bibfnamefont {J.}~\bibnamefont {Kishine}},\ }\bibfield  {title} {\bibinfo {title} {Microscopic {{Theory}} of {{Chiral-Phonon-Induced Orbital Selectivity}} in {{Helical Crystals}}},\ }\href {https://doi.org/10.7566/JPSJ.95.063705} {\bibfield  {journal} {\bibinfo  {journal} {J. Phys. Soc. Jpn.}\ }\textbf {\bibinfo {volume} {95}},\ \bibinfo {pages} {063705} (\bibinfo {year} {2026})}\BibitemShut {NoStop}%
\bibitem [{\citenamefont {Hu}\ \emph {et~al.}(2024)\citenamefont {Hu}, \citenamefont {Zhao}, \citenamefont {Li},\ and\ \citenamefont {Wang}}]{hu2024Electronic}%
  \BibitemOpen
  \bibfield  {author} {\bibinfo {author} {\bibfnamefont {J.}~\bibnamefont {Hu}}, \bibinfo {author} {\bibfnamefont {S.}~\bibnamefont {Zhao}}, \bibinfo {author} {\bibfnamefont {W.}~\bibnamefont {Li}},\ and\ \bibinfo {author} {\bibfnamefont {H.}~\bibnamefont {Wang}},\ }\bibfield  {title} {\bibinfo {title} {Electronic states in one-dimensional helical crystals: {{General}} properties and application to {{InSeI}}},\ }\href {https://doi.org/10.1103/PhysRevB.109.195160} {\bibfield  {journal} {\bibinfo  {journal} {Phys. Rev. B}\ }\textbf {\bibinfo {volume} {109}},\ \bibinfo {pages} {195160} (\bibinfo {year} {2024})}\BibitemShut {NoStop}%
\bibitem [{\citenamefont {R{\"o}ssler}(2009)}]{rossler2009Solid}%
  \BibitemOpen
  \bibfield  {author} {\bibinfo {author} {\bibfnamefont {U.}~\bibnamefont {R{\"o}ssler}},\ }\href@noop {} {\emph {\bibinfo {title} {Solid State Theory: An Introduction}}},\ \bibinfo {edition} {2nd}\ ed.\ (\bibinfo  {publisher} {Springer},\ \bibinfo {address} {Berlin, New York},\ \bibinfo {year} {2009})\BibitemShut {NoStop}%
\bibitem [{\citenamefont {Schrieffer}\ and\ \citenamefont {Wolff}(1966)}]{schrieffer1966Relation}%
  \BibitemOpen
  \bibfield  {author} {\bibinfo {author} {\bibfnamefont {J.~R.}\ \bibnamefont {Schrieffer}}\ and\ \bibinfo {author} {\bibfnamefont {P.~A.}\ \bibnamefont {Wolff}},\ }\bibfield  {title} {\bibinfo {title} {Relation between the {{Anderson}} and {{Kondo Hamiltonians}}},\ }\href {https://doi.org/10.1103/PhysRev.149.491} {\bibfield  {journal} {\bibinfo  {journal} {Phys. Rev.}\ }\textbf {\bibinfo {volume} {149}},\ \bibinfo {pages} {491} (\bibinfo {year} {1966})}\BibitemShut {NoStop}%
\bibitem [{\citenamefont {Winkler}(2003)}]{winkler2003Spin}%
  \BibitemOpen
  \bibfield  {author} {\bibinfo {author} {\bibfnamefont {R.}~\bibnamefont {Winkler}},\ }\href {https://doi.org/10.1007/b13586} {\emph {\bibinfo {title} {Spin---{{Orbit Coupling Effects}} in {{Two-Dimensional Electron}} and {{Hole Systems}}}}},\ \bibinfo {series} {Springer {{Tracts}} in {{Modern Physics}}}, Vol.\ \bibinfo {volume} {191}\ (\bibinfo  {publisher} {Springer},\ \bibinfo {address} {Berlin, Heidelberg},\ \bibinfo {year} {2003})\BibitemShut {NoStop}%
\bibitem [{\citenamefont {Bozovic}\ \emph {et~al.}(1978)\citenamefont {Bozovic}, \citenamefont {Vujicic},\ and\ \citenamefont {Herbut}}]{bozovic1978Irreducible}%
  \BibitemOpen
  \bibfield  {author} {\bibinfo {author} {\bibfnamefont {I.~B.}\ \bibnamefont {Bozovic}}, \bibinfo {author} {\bibfnamefont {M.}~\bibnamefont {Vujicic}},\ and\ \bibinfo {author} {\bibfnamefont {F.}~\bibnamefont {Herbut}},\ }\bibfield  {title} {\bibinfo {title} {Irreducible representations of the symmetry groups of polymer molecules. {{I}}},\ }\href {https://doi.org/10.1088/0305-4470/11/11/003} {\bibfield  {journal} {\bibinfo  {journal} {J. Phys. A: Math. Gen.}\ }\textbf {\bibinfo {volume} {11}},\ \bibinfo {pages} {2133} (\bibinfo {year} {1978})}\BibitemShut {NoStop}%
\bibitem [{\citenamefont {Juraschek}\ \emph {et~al.}(2025)\citenamefont {Juraschek}, \citenamefont {Geilhufe}, \citenamefont {Zhu}, \citenamefont {Basini}, \citenamefont {Baum}, \citenamefont {Baydin}, \citenamefont {Chaudhary}, \citenamefont {Fechner}, \citenamefont {Flebus}, \citenamefont {Grissonnanche}, \citenamefont {Kirilyuk}, \citenamefont {Lemeshko}, \citenamefont {Maehrlein}, \citenamefont {Mignolet}, \citenamefont {Murakami}, \citenamefont {Niu}, \citenamefont {Nowak}, \citenamefont {Romao}, \citenamefont {Rostami}, \citenamefont {Satoh}, \citenamefont {Spaldin}, \citenamefont {Ueda},\ and\ \citenamefont {Zhang}}]{juraschek2025Chiral}%
  \BibitemOpen
  \bibfield  {author} {\bibinfo {author} {\bibfnamefont {D.~M.}\ \bibnamefont {Juraschek}}, \bibinfo {author} {\bibfnamefont {R.~M.}\ \bibnamefont {Geilhufe}}, \bibinfo {author} {\bibfnamefont {H.}~\bibnamefont {Zhu}}, \bibinfo {author} {\bibfnamefont {M.}~\bibnamefont {Basini}}, \bibinfo {author} {\bibfnamefont {P.}~\bibnamefont {Baum}}, \bibinfo {author} {\bibfnamefont {A.}~\bibnamefont {Baydin}}, \bibinfo {author} {\bibfnamefont {S.}~\bibnamefont {Chaudhary}}, \bibinfo {author} {\bibfnamefont {M.}~\bibnamefont {Fechner}}, \bibinfo {author} {\bibfnamefont {B.}~\bibnamefont {Flebus}}, \bibinfo {author} {\bibfnamefont {G.}~\bibnamefont {Grissonnanche}}, \bibinfo {author} {\bibfnamefont {A.~I.}\ \bibnamefont {Kirilyuk}}, \bibinfo {author} {\bibfnamefont {M.}~\bibnamefont {Lemeshko}}, \bibinfo {author} {\bibfnamefont {S.~F.}\ \bibnamefont {Maehrlein}}, \bibinfo {author} {\bibfnamefont {M.}~\bibnamefont {Mignolet}}, \bibinfo {author} {\bibfnamefont {S.}~\bibnamefont {Murakami}}, \bibinfo {author} {\bibfnamefont {Q.}~\bibnamefont {Niu}}, \bibinfo {author} {\bibfnamefont {U.}~\bibnamefont {Nowak}}, \bibinfo {author} {\bibfnamefont {C.~P.}\ \bibnamefont {Romao}}, \bibinfo {author} {\bibfnamefont {H.}~\bibnamefont {Rostami}}, \bibinfo {author} {\bibfnamefont {T.}~\bibnamefont {Satoh}}, \bibinfo {author} {\bibfnamefont {N.~A.}\ \bibnamefont {Spaldin}}, \bibinfo {author} {\bibfnamefont {H.}~\bibnamefont {Ueda}},\ and\ \bibinfo {author} {\bibfnamefont {L.}~\bibnamefont {Zhang}},\ }\bibfield  {title} {\bibinfo {title} {Chiral phonons},\ }\href {https://doi.org/10.1038/s41567-025-03001-9} {\bibfield  {journal} {\bibinfo  {journal} {Nat. Phys.}\ }\textbf {\bibinfo {volume} {21}},\ \bibinfo {pages} {1532} (\bibinfo {year} {2025})}\BibitemShut {NoStop}%
\bibitem [{\citenamefont {Xiao}\ \emph {et~al.}(2010)\citenamefont {Xiao}, \citenamefont {Chang},\ and\ \citenamefont {Niu}}]{xiao2010Berry}%
  \BibitemOpen
  \bibfield  {author} {\bibinfo {author} {\bibfnamefont {D.}~\bibnamefont {Xiao}}, \bibinfo {author} {\bibfnamefont {M.-C.}\ \bibnamefont {Chang}},\ and\ \bibinfo {author} {\bibfnamefont {Q.}~\bibnamefont {Niu}},\ }\bibfield  {title} {\bibinfo {title} {Berry phase effects on electronic properties},\ }\href {https://doi.org/10.1103/RevModPhys.82.1959} {\bibfield  {journal} {\bibinfo  {journal} {Rev. Mod. Phys.}\ }\textbf {\bibinfo {volume} {82}},\ \bibinfo {pages} {1959} (\bibinfo {year} {2010})}\BibitemShut {NoStop}%
\bibitem [{\citenamefont {Pezo}\ \emph {et~al.}(2022)\citenamefont {Pezo}, \citenamefont {Garc{\'i}a~Ovalle},\ and\ \citenamefont {Manchon}}]{pezo2022Orbital}%
  \BibitemOpen
  \bibfield  {author} {\bibinfo {author} {\bibfnamefont {A.}~\bibnamefont {Pezo}}, \bibinfo {author} {\bibfnamefont {D.}~\bibnamefont {Garc{\'i}a~Ovalle}},\ and\ \bibinfo {author} {\bibfnamefont {A.}~\bibnamefont {Manchon}},\ }\bibfield  {title} {\bibinfo {title} {Orbital {{Hall}} effect in crystals: {{Interatomic}} versus intra-atomic contributions},\ }\href {https://doi.org/10.1103/PhysRevB.106.104414} {\bibfield  {journal} {\bibinfo  {journal} {Phys. Rev. B}\ }\textbf {\bibinfo {volume} {106}},\ \bibinfo {pages} {104414} (\bibinfo {year} {2022})}\BibitemShut {NoStop}%
\bibitem [{\citenamefont {Burgos~Atencia}\ \emph {et~al.}(2024)\citenamefont {Burgos~Atencia}, \citenamefont {Agarwal},\ and\ \citenamefont {Culcer}}]{burgosatencia2024Orbital}%
  \BibitemOpen
  \bibfield  {author} {\bibinfo {author} {\bibfnamefont {R.}~\bibnamefont {Burgos~Atencia}}, \bibinfo {author} {\bibfnamefont {A.}~\bibnamefont {Agarwal}},\ and\ \bibinfo {author} {\bibfnamefont {D.}~\bibnamefont {Culcer}},\ }\bibfield  {title} {\bibinfo {title} {Orbital angular momentum of {{Bloch}} electrons: Equilibrium formulation, magneto-electric phenomena, and the orbital {{Hall}} effect},\ }\href {https://doi.org/10.1080/23746149.2024.2371972} {\bibfield  {journal} {\bibinfo  {journal} {Adv. Phys. X}\ }\textbf {\bibinfo {volume} {9}},\ \bibinfo {pages} {2371972} (\bibinfo {year} {2024})}\BibitemShut {NoStop}%
\bibitem [{\citenamefont {Luo}\ \emph {et~al.}(2023)\citenamefont {Luo}, \citenamefont {Lin}, \citenamefont {Zhang}, \citenamefont {Chen}, \citenamefont {Blackert}, \citenamefont {Xu}, \citenamefont {Yakobson},\ and\ \citenamefont {Zhu}}]{luo2023Large}%
  \BibitemOpen
  \bibfield  {author} {\bibinfo {author} {\bibfnamefont {J.}~\bibnamefont {Luo}}, \bibinfo {author} {\bibfnamefont {T.}~\bibnamefont {Lin}}, \bibinfo {author} {\bibfnamefont {J.}~\bibnamefont {Zhang}}, \bibinfo {author} {\bibfnamefont {X.}~\bibnamefont {Chen}}, \bibinfo {author} {\bibfnamefont {E.~R.}\ \bibnamefont {Blackert}}, \bibinfo {author} {\bibfnamefont {R.}~\bibnamefont {Xu}}, \bibinfo {author} {\bibfnamefont {B.~I.}\ \bibnamefont {Yakobson}},\ and\ \bibinfo {author} {\bibfnamefont {H.}~\bibnamefont {Zhu}},\ }\bibfield  {title} {\bibinfo {title} {Large effective magnetic fields from chiral phonons in rare-earth halides},\ }\href {https://doi.org/10.1126/science.adi9601} {\bibfield  {journal} {\bibinfo  {journal} {Science}\ }\textbf {\bibinfo {volume} {382}},\ \bibinfo {pages} {698} (\bibinfo {year} {2023})}\BibitemShut {NoStop}%
\bibitem [{\citenamefont {Basini}\ \emph {et~al.}(2024)\citenamefont {Basini}, \citenamefont {Pancaldi}, \citenamefont {Wehinger}, \citenamefont {Udina}, \citenamefont {Unikandanunni}, \citenamefont {Tadano}, \citenamefont {Hoffmann}, \citenamefont {Balatsky},\ and\ \citenamefont {Bonetti}}]{basini2024Terahertz}%
  \BibitemOpen
  \bibfield  {author} {\bibinfo {author} {\bibfnamefont {M.}~\bibnamefont {Basini}}, \bibinfo {author} {\bibfnamefont {M.}~\bibnamefont {Pancaldi}}, \bibinfo {author} {\bibfnamefont {B.}~\bibnamefont {Wehinger}}, \bibinfo {author} {\bibfnamefont {M.}~\bibnamefont {Udina}}, \bibinfo {author} {\bibfnamefont {V.}~\bibnamefont {Unikandanunni}}, \bibinfo {author} {\bibfnamefont {T.}~\bibnamefont {Tadano}}, \bibinfo {author} {\bibfnamefont {M.~C.}\ \bibnamefont {Hoffmann}}, \bibinfo {author} {\bibfnamefont {A.~V.}\ \bibnamefont {Balatsky}},\ and\ \bibinfo {author} {\bibfnamefont {S.}~\bibnamefont {Bonetti}},\ }\bibfield  {title} {\bibinfo {title} {Terahertz electric-field-driven dynamical multiferroicity in {{SrTiO3}}},\ }\href {https://doi.org/10.1038/s41586-024-07175-9} {\bibfield  {journal} {\bibinfo  {journal} {Nature}\ }\textbf {\bibinfo {volume} {628}},\ \bibinfo {pages} {534} (\bibinfo {year} {2024})}\BibitemShut {NoStop}%
\bibitem [{\citenamefont {Davies}\ \emph {et~al.}(2024)\citenamefont {Davies}, \citenamefont {Fennema}, \citenamefont {Tsukamoto}, \citenamefont {Razdolski}, \citenamefont {Kimel},\ and\ \citenamefont {Kirilyuk}}]{davies2024Phononic}%
  \BibitemOpen
  \bibfield  {author} {\bibinfo {author} {\bibfnamefont {C.~S.}\ \bibnamefont {Davies}}, \bibinfo {author} {\bibfnamefont {F.~G.~N.}\ \bibnamefont {Fennema}}, \bibinfo {author} {\bibfnamefont {A.}~\bibnamefont {Tsukamoto}}, \bibinfo {author} {\bibfnamefont {I.}~\bibnamefont {Razdolski}}, \bibinfo {author} {\bibfnamefont {A.~V.}\ \bibnamefont {Kimel}},\ and\ \bibinfo {author} {\bibfnamefont {A.}~\bibnamefont {Kirilyuk}},\ }\bibfield  {title} {\bibinfo {title} {Phononic switching of magnetization by the ultrafast {{Barnett}} effect},\ }\href {https://doi.org/10.1038/s41586-024-07200-x} {\bibfield  {journal} {\bibinfo  {journal} {Nature}\ }\textbf {\bibinfo {volume} {628}},\ \bibinfo {pages} {540} (\bibinfo {year} {2024})}\BibitemShut {NoStop}%
\bibitem [{\citenamefont {Liu}\ \emph {et~al.}(2021)\citenamefont {Liu}, \citenamefont {Xiao}, \citenamefont {Koo},\ and\ \citenamefont {Yan}}]{liu2021Chiralitydriven}%
  \BibitemOpen
  \bibfield  {author} {\bibinfo {author} {\bibfnamefont {Y.}~\bibnamefont {Liu}}, \bibinfo {author} {\bibfnamefont {J.}~\bibnamefont {Xiao}}, \bibinfo {author} {\bibfnamefont {J.}~\bibnamefont {Koo}},\ and\ \bibinfo {author} {\bibfnamefont {B.}~\bibnamefont {Yan}},\ }\bibfield  {title} {\bibinfo {title} {Chirality-driven topological electronic structure of {{DNA-like}} materials},\ }\href {https://doi.org/10.1038/s41563-021-00924-5} {\bibfield  {journal} {\bibinfo  {journal} {Nat. Mater.}\ }\textbf {\bibinfo {volume} {20}},\ \bibinfo {pages} {638} (\bibinfo {year} {2021})}\BibitemShut {NoStop}%
\bibitem [{\citenamefont {Cho}\ \emph {et~al.}(2026)\citenamefont {Cho}, \citenamefont {Lim},\ and\ \citenamefont {Plenio}}]{cho2026ChiralityInduced}%
  \BibitemOpen
  \bibfield  {author} {\bibinfo {author} {\bibfnamefont {N.}~\bibnamefont {Cho}}, \bibinfo {author} {\bibfnamefont {J.}~\bibnamefont {Lim}},\ and\ \bibinfo {author} {\bibfnamefont {M.~B.}\ \bibnamefont {Plenio}},\ }\bibfield  {title} {\bibinfo {title} {Chirality-{{Induced Orbital Selectivity}} through {{Linear-Orbital Coupling}}},\ }\href {https://doi.org/10.1021/acs.jpclett.6c01552} {\bibfield  {journal} {\bibinfo  {journal} {J. Phys. Chem. Lett.}\ }\textbf {\bibinfo {volume} {17}},\ \bibinfo {pages} {8535} (\bibinfo {year} {2026})}\BibitemShut {NoStop}%
\bibitem [{\citenamefont {Sato}\ \emph {et~al.}(2026)\citenamefont {Sato}, \citenamefont {Kato},\ and\ \citenamefont {Manchon}}]{sato2026Orbital}%
  \BibitemOpen
  \bibfield  {author} {\bibinfo {author} {\bibfnamefont {T.}~\bibnamefont {Sato}}, \bibinfo {author} {\bibfnamefont {T.}~\bibnamefont {Kato}},\ and\ \bibinfo {author} {\bibfnamefont {A.}~\bibnamefont {Manchon}},\ }\bibfield  {title} {\bibinfo {title} {Orbital {{Accumulation Induced}} by {{Chiral Phonons}}},\ }\href {https://doi.org/10.1103/2yqn-cyzg} {\bibfield  {journal} {\bibinfo  {journal} {Phys. Rev. Lett.}\ }\textbf {\bibinfo {volume} {137}},\ \bibinfo {pages} {066302} (\bibinfo {year} {2026})}\BibitemShut {NoStop}%
\bibitem [{\citenamefont {Yao}\ \emph {et~al.}(2026)\citenamefont {Yao}, \citenamefont {Go}, \citenamefont {Mokrousov},\ and\ \citenamefont {Murakami}}]{yao2026Dynamical}%
  \BibitemOpen
  \bibfield  {author} {\bibinfo {author} {\bibfnamefont {D.}~\bibnamefont {Yao}}, \bibinfo {author} {\bibfnamefont {D.}~\bibnamefont {Go}}, \bibinfo {author} {\bibfnamefont {Y.}~\bibnamefont {Mokrousov}},\ and\ \bibinfo {author} {\bibfnamefont {S.}~\bibnamefont {Murakami}},\ }\bibfield  {title} {\bibinfo {title} {Dynamical {{Orbital Angular Momentum Induced}} by {{Circularly Polarized Phonons}}},\ }\href {https://doi.org/10.1103/mnth-3g9n} {\bibfield  {journal} {\bibinfo  {journal} {Phys. Rev. Lett.}\ }\textbf {\bibinfo {volume} {137}},\ \bibinfo {pages} {066301} (\bibinfo {year} {2026})}\BibitemShut {NoStop}%
\bibitem [{\citenamefont {Nabei}\ \emph {et~al.}(2026)\citenamefont {Nabei}, \citenamefont {Yang}, \citenamefont {Sun}, \citenamefont {Jones}, \citenamefont {Mai}, \citenamefont {Wang}, \citenamefont {Bodin}, \citenamefont {Pandey}, \citenamefont {Wang}, \citenamefont {Xiong}, \citenamefont {Comstock}, \citenamefont {Ewing}, \citenamefont {Bingen}, \citenamefont {Sun}, \citenamefont {Smirnov}, \citenamefont {Zhang}, \citenamefont {Hoffmann}, \citenamefont {Rao}, \citenamefont {Hu}, \citenamefont {Vardeny}, \citenamefont {Yan}, \citenamefont {Li}, \citenamefont {Zhou}, \citenamefont {Liu},\ and\ \citenamefont {Sun}}]{nabei2026Orbital}%
  \BibitemOpen
  \bibfield  {author} {\bibinfo {author} {\bibfnamefont {Y.}~\bibnamefont {Nabei}}, \bibinfo {author} {\bibfnamefont {C.}~\bibnamefont {Yang}}, \bibinfo {author} {\bibfnamefont {H.}~\bibnamefont {Sun}}, \bibinfo {author} {\bibfnamefont {H.}~\bibnamefont {Jones}}, \bibinfo {author} {\bibfnamefont {T.}~\bibnamefont {Mai}}, \bibinfo {author} {\bibfnamefont {T.}~\bibnamefont {Wang}}, \bibinfo {author} {\bibfnamefont {R.}~\bibnamefont {Bodin}}, \bibinfo {author} {\bibfnamefont {B.}~\bibnamefont {Pandey}}, \bibinfo {author} {\bibfnamefont {Z.}~\bibnamefont {Wang}}, \bibinfo {author} {\bibfnamefont {Y.}~\bibnamefont {Xiong}}, \bibinfo {author} {\bibfnamefont {A.~H.}\ \bibnamefont {Comstock}}, \bibinfo {author} {\bibfnamefont {B.}~\bibnamefont {Ewing}}, \bibinfo {author} {\bibfnamefont {J.}~\bibnamefont {Bingen}}, \bibinfo {author} {\bibfnamefont {R.}~\bibnamefont {Sun}}, \bibinfo {author} {\bibfnamefont {D.}~\bibnamefont {Smirnov}}, \bibinfo {author} {\bibfnamefont {W.}~\bibnamefont {Zhang}}, \bibinfo {author} {\bibfnamefont {A.}~\bibnamefont {Hoffmann}}, \bibinfo {author} {\bibfnamefont {R.}~\bibnamefont {Rao}}, \bibinfo {author} {\bibfnamefont {M.}~\bibnamefont {Hu}}, \bibinfo {author} {\bibfnamefont {Z.~V.}\ \bibnamefont {Vardeny}}, \bibinfo {author} {\bibfnamefont {B.}~\bibnamefont {Yan}}, \bibinfo {author} {\bibfnamefont {X.}~\bibnamefont {Li}}, \bibinfo {author} {\bibfnamefont {J.}~\bibnamefont {Zhou}}, \bibinfo {author} {\bibfnamefont {J.}~\bibnamefont {Liu}},\ and\ \bibinfo {author} {\bibfnamefont {D.}~\bibnamefont {Sun}},\ }\bibfield  {title} {\bibinfo {title} {Orbital {{Seebeck}} effect induced by chiral phonons},\ }\href {https://doi.org/10.1038/s41567-025-03134-x} {\bibfield  {journal} {\bibinfo  {journal} {Nat. Phys.}\ }\textbf {\bibinfo {volume} {22}},\ \bibinfo {pages} {245} (\bibinfo {year} {2026})}\BibitemShut {NoStop}%
\bibitem [{\citenamefont {Kim}\ \emph {et~al.}(2023)\citenamefont {Kim}, \citenamefont {Vetter}, \citenamefont {Yan}, \citenamefont {Yang}, \citenamefont {Wang}, \citenamefont {Sun}, \citenamefont {Yang}, \citenamefont {Comstock}, \citenamefont {Li}, \citenamefont {Zhou}, \citenamefont {Zhang}, \citenamefont {You}, \citenamefont {Sun},\ and\ \citenamefont {Liu}}]{kim2023Chiralphononactivated}%
  \BibitemOpen
  \bibfield  {author} {\bibinfo {author} {\bibfnamefont {K.}~\bibnamefont {Kim}}, \bibinfo {author} {\bibfnamefont {E.}~\bibnamefont {Vetter}}, \bibinfo {author} {\bibfnamefont {L.}~\bibnamefont {Yan}}, \bibinfo {author} {\bibfnamefont {C.}~\bibnamefont {Yang}}, \bibinfo {author} {\bibfnamefont {Z.}~\bibnamefont {Wang}}, \bibinfo {author} {\bibfnamefont {R.}~\bibnamefont {Sun}}, \bibinfo {author} {\bibfnamefont {Y.}~\bibnamefont {Yang}}, \bibinfo {author} {\bibfnamefont {A.~H.}\ \bibnamefont {Comstock}}, \bibinfo {author} {\bibfnamefont {X.}~\bibnamefont {Li}}, \bibinfo {author} {\bibfnamefont {J.}~\bibnamefont {Zhou}}, \bibinfo {author} {\bibfnamefont {L.}~\bibnamefont {Zhang}}, \bibinfo {author} {\bibfnamefont {W.}~\bibnamefont {You}}, \bibinfo {author} {\bibfnamefont {D.}~\bibnamefont {Sun}},\ and\ \bibinfo {author} {\bibfnamefont {J.}~\bibnamefont {Liu}},\ }\bibfield  {title} {\bibinfo {title} {Chiral-phonon-activated spin {{Seebeck}} effect},\ }\href {https://doi.org/10.1038/s41563-023-01473-9} {\bibfield  {journal} {\bibinfo  {journal} {Nat. Mater.}\ }\textbf {\bibinfo {volume} {22}},\ \bibinfo {pages} {322} (\bibinfo {year} {2023})}\BibitemShut {NoStop}%
\bibitem [{\citenamefont {Bari{\v s}i{\'c}}\ \emph {et~al.}(1970)\citenamefont {Bari{\v s}i{\'c}}, \citenamefont {Labb{\'e}},\ and\ \citenamefont {Friedel}}]{Friedel1970}%
  \BibitemOpen
  \bibfield  {author} {\bibinfo {author} {\bibfnamefont {S.}~\bibnamefont {Bari{\v s}i{\'c}}}, \bibinfo {author} {\bibfnamefont {J.}~\bibnamefont {Labb{\'e}}},\ and\ \bibinfo {author} {\bibfnamefont {J.}~\bibnamefont {Friedel}},\ }\bibfield  {title} {\bibinfo {title} {Tight {{Binding}} and {{Transition-Metal Superconductivity}}},\ }\href {https://doi.org/10.1103/PhysRevLett.25.919} {\bibfield  {journal} {\bibinfo  {journal} {Phys. Rev. Lett.}\ }\textbf {\bibinfo {volume} {25}},\ \bibinfo {pages} {919} (\bibinfo {year} {1970})}\BibitemShut {NoStop}%
\end{thebibliography}%

\end{document}